\documentclass[manuscript,screen]{acmart}

\AtBeginDocument{%
  \providecommand\BibTeX{{%
    \normalfont B\kern-0.5em{\scshape i\kern-0.25em b}\kern-0.8em\TeX}}}

\usepackage{threeparttable}
\usepackage{amsmath}
\usepackage{colortbl}
\usepackage{xcolor}
\usepackage{tabularray}
\usepackage{longtable}
\usepackage{tcolorbox}
\tcbuselibrary{breakable}
\usepackage{multirow}
\usepackage{cite}
\usepackage{hyperref}
\usepackage{makecell}
\usepackage{tocloft}
\usepackage{enumitem}
\usepackage{natbib}
\usepackage{array}
\usepackage{enumitem}
\usepackage{makecell}
\usepackage[title]{appendix}
\newcommand{\tabincell}[2]{\begin{tabular}{@{}#1@{}}#2\end{tabular}}

\setcopyright{acmcopyright}
\copyrightyear{2018}
\acmYear{2018}
\acmDOI{XXXXXXX.XXXXXXX}

\begin{document}
\title{
From 5G to 6G: A Survey on Security, Privacy, and Standardization Pathways
}

\author{Mengmeng Yang}
\author{Youyang Qu}
\author{Thilina Ranbaduge}
\author{Chandra Thapa}
\author{Nazatul Sultan}
\author{Ming Ding}
\author{Hajime Suzuki}
\author{Wei Ni}
\authornote{Corresponding author.}
\email{wei.ni@data61.csiro.au}
\author{Sharif Abuadbba}
\author{David Smith}
\author{Paul Tyler}
\author{Josef Pieprzyk}
\author{Thierry Rakotoarivelo}
\author{Xinlong Guan}
\author{Sirine M'rabet}

\affiliation{%
  \institution{Data61, CSIRO}
  \streetaddress{13 Garden St}
  \city{Sydney}
  \state{NSW}
  \country{Australia}
  \postcode{2015}
}








 \renewcommand{\shortauthors}{Mengmeng Yang and Youyang Qu, et al.}

\begin{abstract}
The vision for 6G aims to enhance network capabilities with faster data rates, near-zero latency, and higher capacity, supporting more connected devices and seamless experiences within an intelligent digital ecosystem where artificial intelligence (AI) plays a crucial role in network management and data analysis. 
This advancement seeks to enable immersive mixed-reality experiences, holographic communications, and smart city infrastructures. 
However, the expansion of 6G raises critical security and privacy concerns, such as unauthorized access and data breaches. 
This is due to the increased integration of IoT devices, edge computing, and AI-driven analytics.
This paper provides a comprehensive overview of 6G protocols, focusing on security and privacy, identifying risks, and presenting mitigation strategies.
The survey examines current risk assessment frameworks and advocates for tailored 6G solutions.
We further discuss industry visions, government projects, and standardization efforts to balance technological innovation with robust security and privacy measures.
\end{abstract}



\received{20 February 2007}
\received[revised]{12 March 2009}
\received[accepted]{5 June 2009}

\maketitle

\section{Introduction}
Wireless communication has undergone rapid evolution, transforming the way we connect and interact. Beginning with basic analog transmissions in the early 20th century, wireless communication evolved into cellular networks, which have since progressed through multiple generations. Each new generation, from 1G's analog cellular networks to 2G's digital signals, and from 3G's introduction of mobile internet to 4G's high-speed broadband, has marked a substantial technological advancement. The advent of 5G further revolutionized wireless communication with its high-speed, low-latency data transfer, enabling real-time applications, the Internet of Things (IoT), and edge computing. Now, 6G emerges as the next frontier.
6G represents a substantial leap from 5G, introducing innovations that extend beyond traditional communication paradigms. 
Beyond technical enhancements, 6G envisions a more interconnected, intelligent digital ecosystem, integrating AI into network management, decision-making, and data analysis. It will support applications, e.g., immersive mixed-reality experiences, holographic communications, and smart cities. 

\begin{figure*}[t]
\centering
\includegraphics[width=15cm]{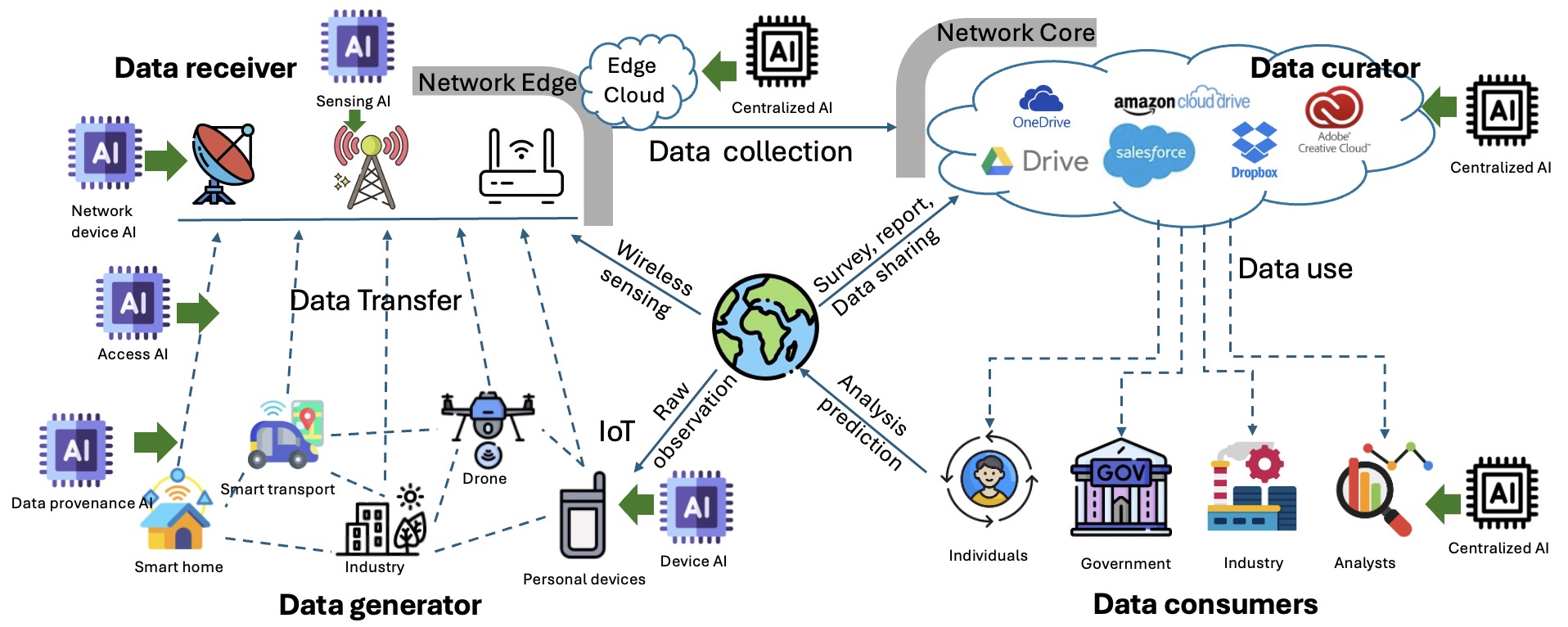}
\caption{Data lifecycle in 6G environments.}
\label{fig:datalifecycle}
\end{figure*}

The substantial increase in coverage and network heterogeneity in 6G introduces severe concerns about security and privacy, potentially exacerbating issues from previous generations. For example, the integration of diverse IoT devices, edge computing, and advanced AI-driven analytics means more data is being collected, processed, and transferred across a broader range of devices and platforms, increasing the risk of unauthorized access, data breaches, and misuse. Fig. \ref{fig:datalifecycle} illustrates a comprehensive data lifecycle in a 6G environment, where AI plays a significant role. The cycle demonstrates how the 6G environment, with its extensive AI integration, significantly accelerates and enhances data collection, transfer, storage, and utilization, making it a powerful ecosystem for data-driven decision-making. At the same time, the integration of AI into the data lifecycle introduces significant privacy and security concerns. As data flows from generators to receivers, curators, and eventually consumers, sensitive information may be exposed at various stages. Trained AI models can become new attack surfaces for security threats where malicious users may target these models, undermining their accuracy~\citep{10518175,10419367,10.1145/3589335.3651490}. The data-driven training process of machine learning (ML) models poses a significant risk to privacy. It may lead to information leakage, malicious tampering, and unauthorized data use. 

\begin{figure*}[!h]
\centering
\includegraphics[width=15cm]{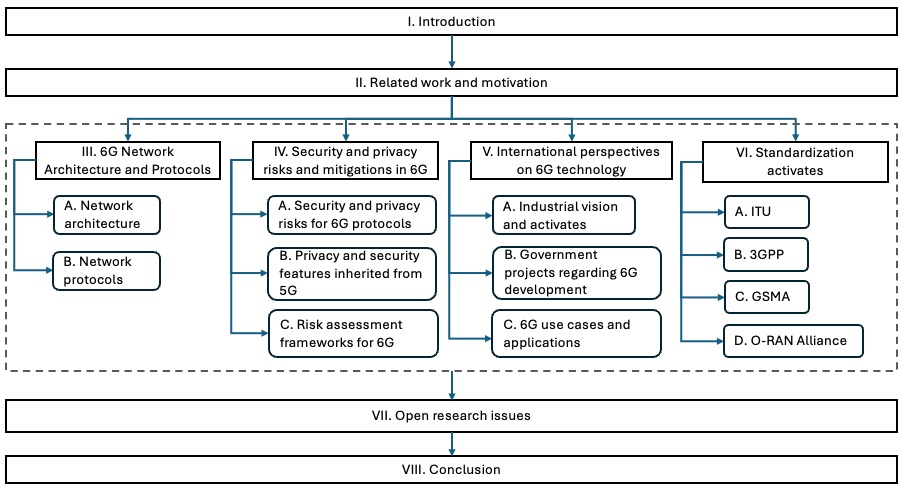}
\caption{The structure of the paper.}
\label{fig:structure}
\end{figure*}

{\color{black}This survey} provides a comprehensive overview of 6G protocols, emphasizing security and privacy concerns while building upon past experiences from earlier generations. It explores the evolution of network architectures, identifies existing and emerging risks associated with 6G protocols, and presents strategies to mitigate these challenges. Additionally, the survey assesses current risk assessment frameworks and their limitations, advocating for tailored solutions for 6G environments. The survey also discusses industry vision, government projects, and standardization efforts that shape 6G's development, balancing innovation with comprehensive security and privacy measures and ensuring consistent deployment and adherence to established standards. Fig. \ref{fig:structure} shows the structure of the paper. 

\textbf{Roadmap.} Section \ref{rwm} discusses similar surveys, highlighting the differences and unique contributions of our work, as well as the motivation behind this survey. Section \ref{nap} presents the 6G architecture and related protocols, detailing the network's evolution and expected advancements. Section \ref{sprm} discusses the security and privacy risks in 6G protocols and solutions to mitigate these risks. The international view of 6G development, encompassing industry vision and government projects, is introduced in Section \ref{ipt}.  Section \ref{sa} covers standardization activities, highlighting how they shape 6G networks. {\color{black}Sections \ref{ori} and \ref{conclusion} discuss the open research issue and conclude the survey, respectively}. Some important acronyms are given in Table \ref{tab:acronyms}. 

\begin{table*}[!htp]
    \centering
    \footnotesize
    \caption{Summary of important acronyms}
    \begin{tabular}{|l|l|l|l|}
    \hline
    3GPP & 3rd Generation Partnership Project & \textcolor{black}{NOMA} & \textcolor{black}{Nonorthogonal Multiple Access} \\ \hline
    AI & Artificial Intelligence & NSA & National Security Agency \\ \hline
    AM & Acknowledged Mode & NTIA & National Telecom and Information Administration \\ \hline
    AMF & Access and Mobility Management Function & O-RAN & Open Radio Access Network \\ \hline
    APT & Advanced Persistent Threat & PCG & Project Co-ordination Group \\ \hline
    AR & Augmented Reality & PDCP & Packet Data Convergence Protocols \\ \hline
    ARQ & Automatic Repeat reQuest & PDN & Packet Data Network \\ \hline
    ATIS & Alliance for Telecommunications Industry Solutions & PGW & Packet Data Network Gateway \\ \hline
    CCAM & Cooperative, Connected, and Automated Mobility & PIoT & Personal IoT \\ \hline
    CISA & Cybersecurity and Infrastructure Security Agency & PLMN & Public Land Mobile Network \\ \hline
    CN & Core Network & PS & Packet-Switch \\ \hline
    CS & Circuit-Switch & PSTN & Public Switched Telephone Network \\ \hline
    DDoS & Distributed Denial of Service & QKD & Quantum Key Distribution \\ \hline
    DLT & Distributed Ledger Technology & QoS & Quality of Service \\ \hline
    DPI & Deep Packet Inspection & RA & Remote Attestation \\ \hline
    DoD & Department of Defense & RAN & Radio Access Network \\ \hline
    EAP & Extensible Authentication Protocol & RCA & Root-cause Analysis \\ \hline
    ENISA & European Union Agency for Cybersecurity & RG & Router Gateway \\ \hline
    EPC & Evolved Packet Core & RIS & Reconfigurable Intelligent Surfaces \\ \hline
    ESF & Enduring Security Framework & RL & Reinforcement Learning \\ \hline
    FDMA & Frequency Division Multiple Access & RLC & Radio Link Control \\ \hline
    FL & Federated Learning & RM & Risk Management \\ \hline
    GDPR & General Data Protection Regulation & RRC & Radio Resource Control \\ \hline
    GSMA & GSM Association & SAE & Security Analytics Engine \\ \hline
    ICS & Information and Communication Systems & SBA & Serviced-based Architecture \\ \hline
    IEEE & Institute of Electrical and Electronics Engineers & SDN & Software Defined Network \\ \hline
    IETF & Internet Engineering TAsk Force & SEPP & Security Edge Protection Proxy \\ \hline
    IMT & International Mobile Telecommunications & \textcolor{black}{SERIMA} & \textcolor{black}{Security Risk Management Platform} \\ \hline
    \textcolor{black}{INSPIRE} & \textcolor{black}{Infrastructure for Spatial Information in the European Community} & SLA & Service Level Agreement \\ \hline
    IPX & IP Exchange & SM & Strategic Measures \\ \hline
    ISP & Internet Service Provider & SMEs & Small and Medium-sized Enterprises \\ \hline
    ITU & international Telecommunication Union & SMF & Session Management Function \\ \hline
    IoE & Internet of Everything & SMS & Short Message Service \\ \hline
    IoT & Internet of Things & SP-MEC & Secure and Private Multi-access Edge Computing \\ \hline
    IoV & Internet of Vehicles & SS7 & Signaling System 7 \\ \hline
    JSON & JavaScript Object Notation & TEE & Trusted Execution Environment  \\ \hline
    JWE & Web Encryption & TM & Transparent Mode \\ \hline
    JWS & JSON Web Signature & TM & Technical Measures \\ \hline
    KEM & Key Encapsulation Mechanism & TNAN & Trusted NOn-3GPP Access Network \\ \hline
    \textcolor{black}{LATERDAF} & \textcolor{black}{Network Data Analytics Function} & TNAP & Trusted Non-3GPP Access Point FUnction \\ \hline
    LIST & Luxemburg Institute of Science and Technology & TNGF & Trusted Non-3GPP Gateway Function \\ \hline
    LPP & LTE Positioning Protocol & TS & Technical Specification \\ \hline
    MCC & Mobile Country Code & TSGs & Technical Specification Groups \\ \hline
    MEC & Multi-access Edge Computing & TWIF & Trusted WLAN Interworking Function \\ \hline
    MIMO & Multiple Input and Multiple Output & UAVs & Unmanned Aerial Vehicles \\ \hline
    ML & Machine Learning & UM & Unacknowledged Mode \\ \hline
    \textcolor{black}{MMT-RCA} & \textcolor{black}{Multi-Modal Traffic Root Cause Analysis} & UMTS & Universal Mobile Telecommunications System \\ \hline
    MNC & Mobile Network Code & UPF & User Plane FUnction \\ \hline
    MNO & Mobile Network Operator & VR & Virtual Reality \\ \hline
    MR & Mixed Reality & \textcolor{black}{VLC} & \textcolor{black}{Visible Light Communication} \\ \hline
    N3IWF & Non-3GPP Internet Function & W-AGF & Wireline Access Gateway Function \\ \hline
    \textcolor{black}{NAD} & \textcolor{black}{Network Anomaly Detection} & WGs & Working Groups \\ \hline
    NAS & Non-Access Stratum & XR & Extended reality \\ \hline
    NF & Network Function & \textcolor{black}{mmWave} & \textcolor{black}{Millimeter Wave} \\ \hline
    NFV & Network Functions Virtualisation & vOBUs & Virtual On-Board Units \\ \hline
    NIST & National Institute of Standards and Technology & & \\ \hline

    \end{tabular}
    \label{tab:acronyms}
\end{table*}

\section{Related survey and motivation} \label{rwm}


\subsection{Existing Surveys}
As summarized in Table \ref{table:similarsurvey}, 
Nguyen et al. \citep{nguyen2021security} and Porambage et al. \citep{porambage2021roadmap} conducted a comprehensive analysis of security and privacy issues in 6G networks. Specifically, Nguyen et al.\citep{nguyen2021security} survey security and privacy challenges for 6G networks, focusing on new threats from advanced technologies like ultra-massive multiple-input multiple-output (MIMO) and pervasive AI. It highlights promising mitigation techniques such as physical layer protection, deep network slicing, quantum-safe communications, and AI-driven security~\citep{10463696}. The paper also explores the necessity for dynamic real-time security and innovative data protection methods such as distributed ledgers and differential privacy, providing a roadmap for future 6G security enhancements. Porambage et al. \citep{porambage2021roadmap} explore future research directions in 6G security and privacy, focusing on trends, applications, requirements, and enabling technologies. It surveys security challenges and potential solutions, including distributed ledger technology (DLT), physical layer security, quantum communication, and distributed AI/ML. Additionally, it presents a roadmap for implementing 6G security visions, highlighting standardization efforts and key research projects driving 6G security advancements. 

All others target one specific domain. Mao et al. \citep{mao2023security} survey security and privacy threats in 6G network edge technologies, focusing on edge computing, caching, and intelligence. 
Ramezanpour et al. \citep{ramezanpour2023security} focus on security and privacy vulnerabilities in 5G/6G and WiFi 6 networks from a coexistence perspective. 
Wang et al. \citep{wang2020security} and Shirwaikar et al. \citep{shirwaikar2021review} focus on four key areas of 6G: real-time intelligent edge computing, distributed AI, intelligent radio, and 3D intercoms. The paper discusses the emerging security and privacy issues associated with these technologies. 
Lu et al. \citep{lu2022reinforcement} survey the application of reinforcement learning (RL) algorithms to enhance security and privacy in 6G networks, addressing vulnerabilities to physical-layer attacks and privacy leakage due to advanced wireless technologies like non-orthogonal multiple-access (NOMA), multi-access edge computing (MEC), millimeter wave (mmWave), massive MIMO, Visible Light Communication (VLC), terahertz (THz), and reconfigurable intelligent surfaces (RIS). 
The study in \citep{ylianttila20206g} identifies key multidisciplinary challenges in creating a trustworthy 6G network, focusing on technology, regulation, economics, politics, and ethics. 
Osorio et al. \citep{osorio2022towards} focus on the security and privacy challenges of integrating 6G technologies into the Internet of Vehicles (IoV). The study highlights the role of promising 6G technologies such as AI, network slicing, blockchain, edge computing, and THz communication in securing IoV. It reviews the security threats and requirements for IoV and provides insights into privacy issues, standardization activities, and potential research directions for secure 6G-enabled IoV systems. 
Naeem et al. \citep{naeem2023security} thoroughly analyze the security and privacy challenges of RIS in 6G networks. It examines the attributes of RIS distinguishing them from other technologies like MIMO and backscatter communication, outlines potential security and privacy attacks in RIS-assisted 6G applications, and explores solutions involving technologies such as mmWave, THz, and device-to-device communication. 
Abdel et al. \citep{abdel2022security} survey the anticipated security and privacy challenges in 6G networks. It highlights emerging technologies like post-quantum cryptography, AI, ML, enhanced edge computing, and blockchain as foundational to 6G. The paper emphasizes the need for novel security methods to meet the advanced requirements of 6G, discussing new approaches for authentication, encryption, access control, and threat detection. 
Je et al. \citep{je2021toward} and Porambage et al. \citep{porambage20216g} analyze new security threats in 6G networks arising from emerging technologies. They discuss the new threats these technologies introduce and outlined possible mitigation strategies to address these evolving threats. 



\begin{table*}
    \centering
    \small
    \caption{Comparison with the existing relevant surveys}
    \label{table:similarsurvey}
    \begin{tabular}{|m{0.6cm}|m{2cm}|m{11cm}|}
    \hline
        \textbf{Ref.} & \textbf{Domain/Focus} &  \textbf{Description} \\
        \hline
        \citep{ramezanpour2023security} & Coexistence& 
       \begin{minipage}[c]{11cm}
        \begin{itemize}[leftmargin=*]
            \item  Focusing on issues arising from their coexistence and spectrum sharing. 
            \item  Highlight new attack vectors and challenges not present in standalone networks, particularly at the physical and medium access control layers. 
           \item  Review potential countermeasures and suggest future research directions to develop security frameworks suitable for these integrated environments.
           \end{itemize}
        \end{minipage}  \\
        \hline 
        \citep{mao2023security} & 6G network edge & 
        \begin{minipage}[c]{11cm}
        \begin{itemize}[leftmargin=*]
            \item Explores the deployment of edge servers and the utilization of 
            techniques like edge computing, edge caching, and edge intelligence 
            for local data storage and processing.
            \item Explores security and privacy risks and discusses various countermeasures. 
            \item Examines the roles of Federated Learning (FL) and blockchain in 
            decentralized edge networks.
        \end{itemize}
        \end{minipage}\\
        \hline
        \citep{naeem2023security} & RIS  & 
        \begin{minipage}[c]{11cm}
        \begin{itemize}[leftmargin=*]
            \item Review the security and privacy challenges associated with the 
            use of RIS in 6G networks. 
            \item  Discusses various security threats and explores potential 
            countermeasures. 
        \end{itemize} 
        \end{minipage}
        \\
     
        \hline
        \citep{osorio2022towards} & IoV & 
        \begin{minipage}[c]{11cm}
        \begin{itemize}[leftmargin=*]
            \item Explores the security and privacy challenges of integrating 6G technologies into IoV.
            \item Discusses how emerging 6G technologies like AI, network softwarization, network slicing, etc. can enhance the security and privacy of IoV systems.
            \item Discusses security threats, requirements, and potential solutions for ensuring secure and reliable IoV.
        \end{itemize} 
        \end{minipage}
        \\
        \hline
        \citep{lu2022reinforcement} & Physical Cross-Layer, RL & 
        \begin{minipage}[c]{11cm}
        \begin{itemize}[leftmargin=*]
            \item Survey the use of RL to enhance physical cross-layer security and 
            privacy in 6G networks. 
            \item Review RL-based solutions to optimize security against smart attacks such as jamming and eavesdropping without relying on precise attack models. 
            \item Explore RL applications in securing unmanned aerial vehicles (UAVs).
        \end{itemize}
        \end{minipage}
        \\
        \hline
        \citep{abdel2022security} & Security and privacy requirements and challenge & 
        \begin{minipage}[c]{11cm}
        \begin{itemize}[leftmargin=*]
            \item Emphasizes the need for novel security methods like new authentication, advanced encryption, access control, and threat detection to meet advanced requirements.
            \item Discusses specific security issues and proposed solutions at the physical layer and AI/ML layers.
            \item Identifies essential security requirements and challenges for key 6G applications, offering a discussion on trustworthiness and potential solutions. 
        \end{itemize} 
        \end{minipage}
        \\
        \hline
        \citep{nguyen2021security} &General 6G & 
        \begin{minipage}[c]{11cm}
        \begin{itemize}[leftmargin=*]
            \item It provides a systematic overview of the security and privacy issues anticipated in 6G networks. 
            \item The paper discusses new threat vectors from emerging radio technologies, and reviews promising techniques for mitigating attacks. 
        \end{itemize}
        \end{minipage}
        \\
        \hline
        \citep{shirwaikar2021review} & Four key areas of 6G &
        \begin{minipage}[c]{11cm}
        \begin{itemize}[leftmargin=*]
            \item Discusses critical technologies of the 6G network: 
            Distributed AI, Real-time Intelligent Edge, Intelligent Radio, and 3D Intercoms.
            \item Investigates the privacy and security challenges associated with these key technologies.
        \end{itemize} 
        \end{minipage}
        \\
        \hline
        \citep{porambage2021roadmap} & General 6G &
        \begin{minipage}[c]{11cm}
        \begin{itemize}[leftmargin=*]
            \item Examines driving trends, visions, applications, requirements, and key enabling technologies related to 6G security and privacy.
            \item Surveys the security solutions for 6G, including distributed ledger technology, physical layer security, quantum communication, and distributed AI/ML.
            \item Introduces standardization initiatives and prominent projects driving the vision for 6G security.
        \end{itemize} 
        \end{minipage}
        \\
        \hline
        \citep{je2021toward} & Emerging technologies & 
        \begin{minipage}[c]{11cm}
        \begin{itemize}[leftmargin=*]
            \item It analyzes new security threats in 6G networks due to emerging technologies like network openness, AI, virtualization, and quantum computing, and presents mitigation strategies.
        \end{itemize} 
        \end{minipage}
        \\
        \hline
        \citep{porambage20216g} & Emerging technologies &\begin{minipage}[c]{11cm}
        \begin{itemize}[leftmargin=*]
            \item Highlights new security and privacy threats from advanced technologies like AI, quantum computing, and distributed ledgers. 
        \end{itemize}
        \end{minipage}
        \\
        \hline
        \citep{wang2020security} & Four key areas of 6G & \begin{minipage}[c]{11cm}
        \begin{itemize}[leftmargin=*]
            \item  Focusing on four key areas: real-time intelligent edge computing, distributed AI, intelligent radio, and 3D intercoms. 
            \item Explore the security challenges associated with emerging 6G technologies. 
            
        \end{itemize} 
        \end{minipage}
        \\
        \hline
        \citep{ylianttila20206g} & Trust in 6G & 
        \begin{minipage}[c]{11cm}
        \begin{itemize}[leftmargin=*]
            \item Identify and address the fundamental research challenges for developing a trustworthy 6G.
            \item  It discusses the necessity of embedded trust in 6G networks, the importance of holistic security architecture, and the complexities of privacy protection in a hyper-connected mobile world.
        \end{itemize} 
        \end{minipage}\\
        \hline
    \end{tabular}
\end{table*}

\subsection{Motivation} 
Existing research has extensively explored various aspects of 6G networks.
However, there is a notable gap in the literature regarding the specific focus on 6G protocols. Addressing the security and privacy issues of 6G protocols is crucial because protocols serve as the backbone of network communication, governing how data is transmitted, processed, and secured. 
In this paper, we provide a comprehensive analysis of the specific security and privacy challenges associated with 6G protocols. This includes a thorough examination of potential vulnerabilities and the development of mitigation strategies to address these risks. We incorporate insights from industry-led projects and governmental initiatives aimed at enhancing the security of 6G protocols. 
We also delve into the ongoing standardization activities related to 6G security and privacy. This includes an examination of the contributions of various standardization bodies and the regulatory frameworks being developed to support secure 6G deployment. 


\section{6G Network Architecture and Protocols}\label{nap}
\subsection{Network architecture} 
The first generation of mobile telecommunication systems was analog-based and mainly introduced voice communication. It used Frequency Division Multiple Access (FDMA) for multiple users to share the available frequency spectrum. 
In 2G and 3G systems, the core network consists of two domains: the circuit-switch (CS) domain and the packet-switch (PS) domain. The CS domain transports phone calls across the geographical region covered by the network operator, communicating with the public switched telephone network (PSTN) and CS domains of other operators. The PS domain handles data streams such as web pages and emails between the user and external packet data networks (PDNs) like the internet. The CS domain uses circuit switching, establishing a dedicated connection for each call, which provides constant data rates with minimal delay but can be inefficient due to over-dimensioning. 
The radio access network (RAN) manages radio communications with the user through two networks: the GSMEDGE radio access network (GERAN) and the UMTS terrestrial radio access network (UTRAN), which use different radio communication techniques but share a common core network. The user's device, known as the user equipment (UE), communicates with the RAN over the air interface, with data flowing downlink from network to UE, and uplink from UE to network.


From the high-level network architecture point of view, the 5G system follows that of the 4G system, but different names are assigned to signify the changes: EPC is replaced by 5G core, and evolved UTRAN (E-UTRAN) is replaced by 5G Radio Access Network (NG-RAN, with ``NG'' stands for next generation). 
The major distinction between LTE and the 5G core is that 5G network functions are designed to be divided into smaller services, allowing them to run in cloud environments like virtual machines and containers. While virtualization is also employed in LTE core network implementations, it was more of an afterthought~\citep{0315}.  
Slicing allows mobile network operators (MNOs) to segment customers into various tenant categories, each with distinct service demands as specified in a Service Level Agreement (SLA)~\citep{10505907}.  
A network slice consistently includes a RAN component and a core network component. This concept is somewhat comparable to DECOR and eDECOR in E-UTRAN, but offers greater flexibility since both 5G Core and NG-RAN were designed from the outset with slicing in consideration \citep{0316}. Another significant difference is that while E-UTRAN supports only one air interface, NG-RAN supports multiple air interfaces, e.g., LTE and 5G air interfaces. 

It is expected that 6G network architecture would follow that of 5G in that it is aimed to be virtualized and cloud-native. 6G network architecture is expected to support the following characteristics/functions \citep{0301}:

\noindent \textbf{Network Programmability}: 6G networks aim to offer advanced programmability, allowing network operators and administrators to dynamically configure, manage, and optimize network functions and services. This would enable real-time adjustments to meet changing user demands, implement new services, and enhance overall network efficiency.

\noindent\textbf{Deployment Flexibility}: 6G networks are designed to offer greater flexibility in deployment.
This flexibility can support diverse applications ranging from urban centers to remote rural areas, and from consumer devices to industrial applications, ensuring that 6G technology can meet a broad range of needs.

\noindent\textbf{Simplicity and Efficiency}: 6G networks are expected to be simpler to manage and more efficient than their predecessors. This includes streamlining network components and reducing complexities in architecture, which can lower operational costs and improve performance. By minimizing redundant processes and making networks more intuitive, 6G aims to deliver enhanced user experiences and reduce maintenance burdens. 

\noindent\textbf{Quantum-Integrated}: 6G will be quantum and post-quantum ready. Quantum computing can solve complex problems at unprecedented speeds, revolutionizing cryptography, optimization, and AI. Quantum communication technologies, like Quantum Key Distribution (QKD), provide highly secure channels. Integrating these into 6G will enhance security and open new possibilities for secure data transfer and connectivity.

\noindent\textbf{Security, Robustness, and Reliability}: 6G networks aim to enhance security measures to protect against evolving threats, ensuring robustness and reliability for critical applications. With the growing reliance on communication networks, 6G architecture will incorporate advanced encryption, intrusion detection, and mitigation strategies to safeguard data, maintain service continuity, and ensure network integrity.

\noindent\textbf{Automation}: 6G will feature extensive automation capabilities, leveraging AI and ML to manage network functions autonomously. It will optimize resource allocation, monitor performance metrics, and anticipate network issues before they arise, leading to reduced downtimes, higher quality of service (QoS), and more efficient network management.

\subsection{Network protocols}
Network protocols are foundational rules and standards that underpin efficient, secure, and reliable communication within and across networks. 
A security protocol serves as a structured communication process between multiple entities to achieve shared security objectives through the application of cryptographic techniques. This includes elements like key establishment, authentication, encryption, secure data transfer, non-repudiation, and more. Such protocols are often defined by standardization bodies to ensure effective and secure communication practices, as seen in mobile broadband standards, such as the 3rd Generation Partnership Project (3GPP) or the Institute of Electrical and Electronics Engineers (IEEE).
6G is still being defined by international standardization bodies, so its exact specifications and protocols are not yet known. Nevertheless, several critical protocols that have been playing some key responsibilities through earlier generations of wireless communication systems, e.g., 4G and 5G, are expected to remain critical in 6G. 

\subsubsection{Legacy Protocols for 6G Operations}

It is likely that 6G will incorporate many of the security measures that have been developed for 5G systems, as well as additional measures explicitly designed to address potential vulnerabilities and threats that may arise with the deployment of 6G technology. 
According to the European Union Agency for Cybersecurity (ENISA) categorization of 5G protocols {\color{black}\citep{kim20205g}}, 6G protocols are also anticipated to include IP and cellular stack.  5G networks support protocols known as Signaling System 7 (SS7) and Diameter to maintain global connectivity (roaming), and so are 6G networks expected to. 
The Non-Access Stratum (NAS) in 4G LTE and 5G networks facilitates control communication between UE and the core network, managing mobility and session functions. The Extensible Authentication Protocol (EAP) is a flexible authentication framework used in networks for secure client-server communication. Radio Resource Control (RRC) manages UE connections with the RAN, handling connection setup, maintenance, and security. The Packet Data Convergence Protocol (PDCP) improves transmission efficiency through IP header compression and encryption. Radio Link Control (RLC) ensures efficient data transfer in 3G, LTE, and 5G networks . The Home/Visitor Public Land Mobile Network (PLMN) manages user connections globally, while the Security Edge Protection Proxy (SEPP) secures communications between network functions of different PLMNs. The Packet Data Network Gateway (PGW) routes data traffic and enforces network policies. The Non-3GPP Interworking Function (N3IWF) enables 5G network integration with non-3GPP networks like Wi-Fi, while the Trusted Non-3GPP Gateway Function (TNGF) provides secure gateways for non-3GPP access. The Wireline Access Gateway Function (W-AGF) connects wireline technologies to the core network, and the Trusted WLAN Interworking Function (TWIF) facilitates secure integration between 5G and WLANs. These components will continue to play crucial roles in 6G.

\subsubsection{Post-quantum Cryptographic Protocols for 6G}
Quantum computers, which are still in their early stages of development, have the potential to break many of the encryption algorithms that are currently in use. This is a concern for the future security of communication systems, including those used in 6G networks.  
It is essential to ensure that these networks are secure and that their encryption algorithms cannot be easily broken. Post-quantum cryptography offers a potential solution to this problem. By using cryptography algorithms that are resistant to attacks by quantum computers, it can provide a high level of security for communication systems that use 6G networks. 

The National Institute of Standards and Technology (NIST) has conducted a competition to select cryptographic algorithms that can withstand quantum attacks. NIST recommends the following post-quantum cryptographic algorithms:
\begin{itemize}[leftmargin=*]  
    \item \textit{Public-Key Encryption and Key Establishment Protocols}: CRYSTALS-KYBER {\color{black} \citep{avanzi2017crystals}}, which secures its operations based on the quantum-resistant hardness of learning with error problem, predominantly lattice-based, is recommended for public-key encryption.
    \item \textit{Digital Signatures}: CRYSTALS-DILITHIUM {\color{black}\citep{lyubashevsky2020crystals}} is selected as the principal option for digital signatures, leveraging lattice-based problems for security. Additionally, FALCON {\color{black}\citep{prest2020falcon}}, known for its shorter signatures and lattice-based framework, and SPHINCS+ {\color{black}\citep{bernstein2019sphincs+}}, a hash-based backup option if lattice problems are compromised, are also endorsed.
\end{itemize}
Migration from traditional cryptographic protocols to those secure against quantum adversaries necessitates changes:
\begin{itemize}[leftmargin=*]  
    \item \textit{Algorithm Selection}: NIST's endorsement includes algorithms like CRYSTALS-KYBER for encryption and CRYSTALS-DILITHIUM for signatures, based on the intractability of lattice problems.

\item \textit{Performance Considerations}: Post-quantum algorithms, while secure, are noted for being slower and requiring longer cryptographic keys compared to their classical counterparts.

\item \textit{Efficiency Issues}: Due to efficiency concerns, public-key encryption is suggested to serve as a key encapsulation mechanism (KEM). This involves a computationally expensive secret key distribution phase using CRYSTALS-KYBER, followed by a more efficient symmetric-key encryption phase using AES {\color{black}\citep{abdullah2017advanced}} with enlarged session keys.

\item \textit{Alternative Quantum Solutions}: The first phase of KEM can be alternatively executed through QKD, although this is limited to setups where appropriate quantum channels, like optical fiber or free air, are available.

\item \textit{Quantum Attacks on Symmetric Keys}: Symmetric-key encryption is vulnerable to Grover's quantum search attack, which reduces the security level by half; for instance, AES-128 {\color{black}\citep{liberatori2007aes}} offers a practical security level akin to 64 bits under quantum conditions.

\item \textit{Cross-Domain Security}: For applications spanning various security domains, such as internet-enabled devices interacting over quantum-vulnerable networks, proxy re-encryption is necessary to elevate classical encryption to quantum-resistant standards.

\item \textit{Implications for Digital Signatures}: Quantum-resistant digital signatures, although more secure, tend to be longer and more computationally intensive compared to classical counterparts like RSA-1024 {\color{black}\citep{shamir2003cost}}.

\item \textit{Quantum Threats to Hashing}: Cryptographic hashing algorithms also face quantum risks, notably from the Brassard-Høyer-Tapp (BHT) attack {\color{black}\citep{bernstein2020discretization}}, which can expedite collision findings significantly faster than classical methods such as the Birthday Paradox {\color{black}\citep{ghosh2012birthday}}.
\end{itemize}

The global response to these quantum threats includes significant investment from major IT players like IBM, Microsoft, and Google, which have rapidly expanding quantum computing research labs. Additionally, there is a burgeoning industry of companies focused solely on quantum technology, such as QuintessenceLabs (Australia), MagiQ Technologies Inc. (USA), ID Quantique (Switzerland), and many others. This expansion is complemented by the development of various quantum networks like the DARPA Quantum Network and the Tokyo Network, which facilitate the secure distribution of secrets among nodes. Collectively, these efforts from different entities highlight a worldwide commitment to advancing quantum technologies and enhancing security against quantum threats. 

\section{Security and privacy risks and mitigation in 6G}\label{sprm}

\subsection{Security and privacy risks for 6G Protocols}
This section explores the vulnerabilities that arise not only from the existing attacks in earlier generations but also from new threats specific to 6G. 

\subsubsection{Existing attacks and risks}

While 6G networks aim to uphold stringent security standards and offer low latency, they remain susceptible to a spectrum of attacks that have plagued previous generations. These attacks may not be newly tailored for 6G, but they persist and adapt within this evolving landscape. 
We group the existing attacks, as shown in Table \ref{EA}. 
Table \ref{tab: risk assessment for legacy protocols} summarizes the risk of the attacks on the 6G protocol interfaces.

\begin{footnotesize}
\begin{longtable}{|m{2cm}|m{3cm}|m{9cm}|}
\caption{Existing attacks} \label{EA}
\\ \hline

   \centering \textbf{Category} &  \centering \textbf{Attacks} & \multicolumn{1}{c|}{ {\footnotesize \textbf{Description}} }\\

    \hline
    \endfirsthead

			\multicolumn{3}{|l|}%
			{{\bfseries  -- continued from previous page}} \\

    \hline
    \textbf{Category} & \textbf{Attacks} & \textbf{Description} \\
    \hline
    \endhead
    
    \hline
       \multicolumn{3}{l}{{Continued on next page}} \\ 
    \endfoot
    \hline
    \endlastfoot
   
    \multirow{4}{*}{\tabincell{c}{Authentication \\failure}} & Parameter attacks & Exploit vulnerabilities in authentication parameters to gain unauthorized access. \\ \cline{2-3}
    & Identity attacks & Impersonate or steal a user's identity to gain access to secure systems. \\ \cline{2-3}
    & Remote attestation & Validates the integrity of remote systems and software to prevent unauthorized access. \\ \cline{2-3}
    & Unauthorized access & Occurs when an attacker gains access to a system or resource without permission. \\ \hline

    \multirow{6}{*}{\tabincell{c}{Communication \\interference}} & Man-in-the-middle attacks & An attacker intercepts and manipulates communication between two parties. \\ \cline{2-3}
    & DoS/DDoS attacks & Floods a network or system with traffic to render it unavailable or disrupt communication. \\ \cline{2-3}
    & Insecure execution platform & A platform that can lead to unauthorized code execution and compromise communication. \\ \cline{2-3}
    & Jamming & Uses electromagnetic interference to disrupt wireless communication signals. \\ \cline{2-3}
    & Deception attacks & Mislead or trick communication participants causing data breaches or mismanagement. \\ \cline{2-3}
    & Information exposure & Unintentional or malicious exposing of sensitive data to unauthorized parties. \\ \hline

    \multirow{3}{*}{\tabincell{c}{Integrity failure}} & Authentication of endpoints & Ensures remote systems are genuine and trustworthy to prevent unauthorized access or manipulation. \\ \cline{2-3}
    & Certification of platforms & Validates and certifies platforms to adherence to standards and avoid malicious software. \\ \cline{2-3}
    & Collusion Attacks & Multiple entities work together to undermine the integrity of a system or communication. \\ \hline

    \multirow{5}{*}{\tabincell{c}{Malicious software}} & Code/SQL injection attacks & Inject malicious code or SQL commands to compromise functionality or data. \\ \cline{2-3}
    & FTCode Ransomware & An attack that encrypts files on a system, demanding payment for their decryption. \\ \cline{2-3}
    & Zero-Day Vulnerabilities, Exploits and Attacks & Exploit previously unknown vulnerabilities in software to execute attacks before patches are available. \\ \cline{2-3}
    & Advanced Persistent Threat (APT) Attacks & Sophisticated, long-term attacks designed to infiltrate and remain undetected within systems for extended periods. \\ \cline{2-3}
    & Zeus Malware & A type of malware designed to steal sensitive information, such as banking credentials, by keylogging and form-grabbing. \\ \hline

    \multirow{9}{*}{\tabincell{c}{Access control \\failure}} & Privilege escalation & An attacker gains higher-level access rights to a system than originally granted. \\ \cline{2-3}
    & Insider threats & Threats from employees who misuse their access to compromise security. \\ \cline{2-3}
    & Password stealing & An attacker steals user passwords to gain unauthorized access to systems or accounts. \\ \cline{2-3}
    & Brute-force attacks & Repeatedly attempts to guess passwords or encryption keys until the correct one is found. \\ \cline{2-3}
    &Credential stuffing & Uses stolen credentials from one system to gain access to other systems where users might reuse passwords. \\ \cline{2-3}
    & Security risks in access and authorization of 6G~subnets & Unauthorized access to 6G sub-networks due to vulnerabilities in access control and authorization mechanisms. \\ \cline{2-3}
    & Undesirable configurations & Insecure system configurations that allow unauthorized access or compromise. \\ \cline{2-3}
    & Abnormal behaviors & Unusual or unexpected behaviors in systems that may indicate a security breach or misuse. \\ \cline{2-3}
    & Malformed intent & Anomalous or maliciously crafted input that leads to unintended or harmful outcomes in systems. \\ \hline

    \multirow{7}{*}{\tabincell{c}{Privacy failure}} & Background Knowledge Attacks & Use external information to infer private data from seemingly anonymous datasets. \\ \cline{2-3}
    & Inference Attacks & Deduce sensitive information from seemingly unrelated data or patterns. \\ \cline{2-3}
    & Linkage Attacks & Combine different datasets to identify individuals or reveal sensitive information. \\ \cline{2-3}
    & Privacy risks in creation/ transmission of digital representations of real/virtual objects & Occurs when sensitive information is unintentionally exposed in the digital representation or transmission process. \\ \cline{2-3}
    & Privacy risks in integration of sensors and comm. devices & Sensitive data may be exposed through unauthorized access to sensors or communication devices. \\ \cline{2-3}
    & Privacy risk in offloading tasks from users to networks & Occurs when sensitive data is exposed during the process of transferring computation from devices to network servers. \\ \cline{2-3}
    & Privacy risks in distributed learning based on user data & ML models trained on user data may reveal sensitive information or insights about individual users. \\ \hline
\end{longtable}

\end{footnotesize}

\begin{figure*}[h]
\begin{small}
\begin{threeparttable}
\centering 
\begin{longtable}{|l|l|l|l|l|l|l|l|l|l|l|l|l|}
\caption{Risk of the attacks for legacy protocols} 
\label{tab: risk assessment for legacy protocols}
\\
            \hline 
		& \multicolumn{1}{l|}{\textbf{NAS}}               
            & \multicolumn{1}{l|}{\textbf{EAP}}               
            & \multicolumn{1}{l|}{\textbf{RRC}}               
            & \multicolumn{1}{l|}{\textbf{\begin{tabular}[c]{@{}l@{}}PD\\ CP\end{tabular}}} 
            & \multicolumn{1}{l|}{\textbf{RLC}}               
            & \multicolumn{1}{l|}{\textbf{\begin{tabular}[c]{@{}l@{}}PL\\ MN\end{tabular}}} 
            & \multicolumn{1}{l|}{\textbf{\begin{tabular}[c]{@{}l@{}}SE\\ PP\end{tabular}}} 
            & \multicolumn{1}{l|}{\textbf{PGW}}               
            & \multicolumn{1}{l|}{\textbf{\begin{tabular}[c]{@{}l@{}}N3I\\ WF\end{tabular}}} 
            & \multicolumn{1}{l|}{\textbf{\begin{tabular}[c]{@{}l@{}}TN\\ GF\end{tabular}}}
            & \multicolumn{1}{l|}{\textbf{\begin{tabular}[c]{@{}l@{}}WA\\ GF\end{tabular}}} 
            & \multicolumn{1}{l|}{\textbf{\begin{tabular}[c]{@{}l@{}}TW\\ IF\end{tabular}}} \\ 
			\endfirsthead
			
			\multicolumn{13}{|l|}%
			{{\bfseries  -- continued from previous page}} \\
			\hline 
            & \multicolumn{1}{l|}{\textbf{NAS}}               
            & \multicolumn{1}{l|}{\textbf{EAP}}               
            & \multicolumn{1}{l|}{\textbf{RRC}}               
            & \multicolumn{1}{l|}{\textbf{\begin{tabular}[c]{@{}l@{}}PD\\ CP\end{tabular}}} 
            & \multicolumn{1}{l|}{\textbf{RLC}}               
            & \multicolumn{1}{l|}{\textbf{\begin{tabular}[c]{@{}l@{}}PL\\ MN\end{tabular}}} 
            & \multicolumn{1}{l|}{\textbf{\begin{tabular}[c]{@{}l@{}}SE\\ PP\end{tabular}}} 
            & \multicolumn{1}{l|}{\textbf{PGW}}               
            & \multicolumn{1}{l|}{\textbf{\begin{tabular}[c]{@{}l@{}}N3I\\ WF\end{tabular}}} 
            & \multicolumn{1}{l|}{\textbf{\begin{tabular}[c]{@{}l@{}}TN\\ GF\end{tabular}}}
            & \multicolumn{1}{l|}{\textbf{\begin{tabular}[c]{@{}l@{}}WA\\ GF\end{tabular}}} 
            & \multicolumn{1}{l|}{\textbf{\begin{tabular}[c]{@{}l@{}}TW\\ IF\end{tabular}}}\\ 
            \hline 
			\endhead
			
			\hline \multicolumn{3}{l}{{Continued on next page}} \\ 
			\endfoot

			\endlastfoot
			\hline

\begin{tabular}[l]{@{}l@{}}Parameter attack\end{tabular} 
&\multicolumn{1}{c|}{\checkmark}  &   &   &&  &   &  &  &    &   &   & 
\\ \hline

\begin{tabular}[l]{@{}l@{}}Identity attack\end{tabular}  &      & \multicolumn{1}{c|}{\checkmark}  
& \multicolumn{1}{c|}{\checkmark}  &  &  &   &   & &  &   &   &   \\ \hline

\begin{tabular}[c]{@{}l@{}}Remote attestation\end{tabular}   &  & \multicolumn{1}{c|}{\checkmark}  
& \multicolumn{1}{c|}{\checkmark}  && & &  &  
& \multicolumn{1}{c|}{\checkmark}                   & \multicolumn{1}{c|}{\checkmark}   &               & \multicolumn{1}{c|}{\checkmark}                   \\ \hline

\begin{tabular}[l]{@{}l@{}}Unauthorized access\end{tabular}   &  & \multicolumn{1}{c|}{\checkmark}  &     &  &       & \multicolumn{1}{c|}{\checkmark}                   & \multicolumn{1}{c|}{\checkmark}   &              & \multicolumn{1}{c|}{\checkmark}    &   &          & \multicolumn{1}{c|}{\checkmark}                   \\ \hline

\begin{tabular}[l]{@{}p{3cm}@{}}Man-in-the middle\end{tabular}
& \multicolumn{1}{c|}{\checkmark}  
& \multicolumn{1}{c|}{\checkmark} 
& \multicolumn{1}{c|}{\checkmark}  
& {\color[HTML]{000000} }                                           & {\color[HTML]{000000} }                         
& \multicolumn{1}{c|}{\checkmark}   &    
& \multicolumn{1}{c|}{\checkmark}  
& \multicolumn{1}{c|}{\checkmark} &     &       
& \multicolumn{1}{c|}{\checkmark}                    \\ \hline

\begin{tabular}[l]{@{}l@{}}DoS/DDOS\end{tabular} 
& \multicolumn{1}{c|}{\checkmark} 
& \multicolumn{1}{c|}{\checkmark}  
& \multicolumn{1}{c|}{\checkmark}  
& \multicolumn{1}{c|}{\checkmark}  
& \multicolumn{1}{c|}{\checkmark}  
& \multicolumn{1}{c|}{\checkmark}                   & \multicolumn{1}{c|}{\checkmark}                   & \multicolumn{1}{c|}{\checkmark}  
& \multicolumn{1}{c|}{\checkmark}                   & \multicolumn{1}{c|}{\checkmark}                   & \multicolumn{1}{c|}{\checkmark}                   & \multicolumn{1}{c|}{\checkmark}   
\\ \hline

\begin{tabular}[l]{@{}l@{}}Insecure  execution  platform\end{tabular}      &   &   
& \multicolumn{1}{c|}{\checkmark}  &    &           & \multicolumn{1}{c|}{\checkmark}     & &           & \multicolumn{1}{c|}{\checkmark}  &  & & \multicolumn{1}{c|}{\checkmark}                                \\ \hline

Jamming    &  &    &     &    &   &  & 
& \multicolumn{1}{c|}{\checkmark}  
& \multicolumn{1}{c|}{\checkmark}    
& \multicolumn{1}{c|}{\checkmark}  &                & \multicolumn{1}{c|}{\checkmark}                   \\ \hline

\begin{tabular}[l]{@{}l@{}}Deception\end{tabular}   &   &   &  &   &  & 
& \multicolumn{1}{c|}{\checkmark}  &   & &   & &     \\ \hline

\begin{tabular}[l]{@{}l@{}}Information exposure\end{tabular}     &   &   &    
& \multicolumn{1}{c|}{\checkmark}  &                & \multicolumn{1}{c|}{\checkmark}    &   &  &   & &   &    \\ \hline

\begin{tabular}[l]{@{}l@{}}Authentication  of remote  endpoints\end{tabular}       &   
& \multicolumn{1}{c|}{\checkmark}  & & &     &      
& \multicolumn{1}{c|}{\checkmark}    &  
& \multicolumn{1}{c|}{\checkmark}  
& \multicolumn{1}{c|}{\checkmark}                   & \multicolumn{1}{c|}{\checkmark}                   & \multicolumn{1}{c|}{\checkmark}                      \\ \hline

\begin{tabular}[l]{@{}l@{}}Certification  of platforms\end{tabular} 
& \multicolumn{1}{c|}{\checkmark}  &   & &          & \multicolumn{1}{c|}{\checkmark}    &    &        & \multicolumn{1}{c|}{\checkmark}    &   &   & &     \\ \hline

\begin{tabular}[l]{@{}l@{}}Collusion  Attacks\end{tabular}   & \multicolumn{1}{c|}{\checkmark}   &  &  & \multicolumn{1}{c|}{\checkmark}    & \multicolumn{1}{c|}{\checkmark}    & \multicolumn{1}{c|}{\checkmark}    &  & &&&&  \\ \hline

\begin{tabular}[c]{@{}l@{}}Code and SQL injection\end{tabular}    &    &  &   &  & &   &  &  & \multicolumn{1}{c|}{\checkmark}    & \multicolumn{1}{c|}{\checkmark}   &    & \multicolumn{1}{c|}{\checkmark}     
\\ \hline

\begin{tabular}[l]{@{}l@{}}FTCode Ransomware\end{tabular}       &  & &   &   &    &     &    
& \multicolumn{1}{c|}{\checkmark}  &    &     &     &   \\ \hline

\begin{tabular}[l]{@{}l@{}}Zero-Day Vulnerabilities\end{tabular}  &   &   &  &    &   &  & 
& \multicolumn{1}{c|}{\checkmark}  &    &     &   &  \\ \hline

\begin{tabular}[l]{@{}l@{}}APT\end{tabular}    &  &   &  
& \multicolumn{1}{c|}{\checkmark}  &    &     &  
& \multicolumn{1}{c|}{\checkmark}  &   &   &    
& \multicolumn{1}{c|}{\checkmark}  
\\ \hline

Zeus Malware    &    & \multicolumn{1}{c|}{\checkmark}  &      &    &   
& \multicolumn{1}{c|}{\checkmark}     
& \multicolumn{1}{c|}{\checkmark}        
& \multicolumn{1}{c|}{\checkmark}  &    &        
& \multicolumn{1}{c|}{\checkmark}     
& \multicolumn{1}{c|}{\checkmark}   
\\ \hline

\begin{tabular}[l]{@{}l@{}}Privilege escalation\end{tabular}    & \multicolumn{1}{c|}{\checkmark}  &    &         
& \multicolumn{1}{c|}{\checkmark}    
& \multicolumn{1}{c|}{\checkmark}  &                & \multicolumn{1}{c|}{\checkmark}                 
& \multicolumn{1}{c|}{\checkmark}  &   &  &  &    \\ \hline

\begin{tabular}[l]{@{}l@{}}Insider threats\end{tabular}  
& \multicolumn{1}{c|}{\checkmark}  &   &    
& \multicolumn{1}{c|}{\checkmark}    &    &     &  &   &     &    &     &   \\ \hline

\begin{tabular}[l]{@{}l@{}}Password Stealing\end{tabular}    &   & \multicolumn{1}{c|}{\checkmark}  &   &   &       & \multicolumn{1}{c|}{\checkmark}   &  &   &  &  &&  \\ \hline

\begin{tabular}[l]{@{}l@{}}Brute-force\end{tabular}  &  & \multicolumn{1}{c|}{\checkmark}  &   &   &    & 
& \multicolumn{1}{c|}{\checkmark}   &   &   &  &&   \\ \hline

\begin{tabular}[l]{@{}l@{}}credential stuffing\end{tabular}  &  &   &   &    &
& \multicolumn{1}{c|}{\checkmark}   &   &   &   &  &   &  \\ \hline

\begin{tabular}[l]{@{}l@{}}Sub-network access control
\end{tabular}&   
& \multicolumn{1}{c|}{\checkmark}  &  & &   & 
& \multicolumn{1}{c|}{\checkmark}  
& \multicolumn{1}{c|}{\checkmark}  
& \multicolumn{1}{c|}{\checkmark}   &    & &    \\ \hline

\begin{tabular}[l]{@{}l@{}}Undesirable  configuration\end{tabular}     
& \multicolumn{1}{c|}{\checkmark}  
& \multicolumn{1}{c|}{\checkmark}  &      &    &   & \multicolumn{1}{c|}{\checkmark}   
& \multicolumn{1}{c|}{\checkmark}        &   
& \multicolumn{1}{c|}{\checkmark}    
& \multicolumn{1}{c|}{\checkmark}  &    
& \multicolumn{1}{c|}{\checkmark}    \\ \hline

\begin{tabular}[l]{@{}l@{}}Abnormal behaviours\end{tabular} 
& \multicolumn{1}{c|}{\checkmark}  &   &  
& \multicolumn{1}{c|}{\checkmark}   
& \multicolumn{1}{c|}{\checkmark}  &  &  
& \multicolumn{1}{c|}{\checkmark}  &    &           & \multicolumn{1}{c|}{\checkmark}      &  \\ \hline

\begin{tabular}[l]{@{}l@{}}Malinformed intent\end{tabular} &&   
& \multicolumn{1}{c|}{\checkmark}  
& \multicolumn{1}{c|}{\checkmark}   
& \multicolumn{1}{c|}{\checkmark}  &   &   & &  &   &   &  \\ \hline			

\end{longtable}
 \footnotesize{NAS stands for ``Non-Access Stratum,'' EAP stands for ``Extensible Authentication Protocol,'' RRC stands for ``Radio Resource Control,'' PDCP stands for ``Packet Data Convergence Protocol,'' RLC stands for ``Radio Link Control,'' PLMN stands for ``Public Land Mobile Network,'' SEPP stands for ``Security Edge Protection Proxy,'' PGW stands for ``Packet data network GateWay,'' N31WF stands for ``Non-3GPP InterWorking," TNGF stands for ``Trusted Non-3GPP Gateway Function,'' WAGF stands for ``Wireline Access Gateway Function,'' and TWIF stands for ``Trusted WLAN InterWorking Function.''}
\end{threeparttable}
\end{small}
\end{figure*}

\subsubsection{Potential vulnerabilities}

The evolution from 5G to 6G is not merely an incremental upgrade but a transformation that introduces profound complexity and novel paradigms in network architecture. This progress, while heralding unprecedented connectivity speeds and data-handling capabilities, also opens new vectors for security threats that must be addressed. Several potential security and privacy challenges may arise from 6G, including:

\begin{itemize}[leftmargin=*]  
    \item 
Increased complexity: As 6G networks will be more complex than 5G networks, it may be more difficult to detect and prevent security threats. 
    \item
Greater use of AI and ML: The increased use of AI and ML in 6G networks may make them more vulnerable to attacks, as these technologies can be used by malicious actors to gather and exploit personal information.
    \item
More distributed networks: As 6G networks will be more distributed, it may be harder to monitor and protect.
    \item
New technologies: Large-scale use of new technologies, e.g., THz communications, holographic communications, and quantum communications, may bring security and privacy threats not considered before.
\end{itemize}

Owing to the novel features of 6G, a range of new security and privacy risks also emerge. These include: 

\noindent\textbf{Network anomaly considering new critical architecture.} 
The growth of IoT has led to a surge of data generated by sensors and smartphones, which are processed in the cloud using AI. As a result, data privacy has become more crucial due to the sensitive client information involved. FL is a distributed ML approach that can train models on end devices without revealing private data.  While preserving privacy, it may rely on unreliable clients with inaccessible datasets. FL can also detect network anomalies and prevent attacks early. 

\noindent\textbf{Insecure and non-private MEC.} 
MEC in 5G/6G provides computing capabilities on the network edge close to end users, reducing latency and core network bandwidth. Optimizing the placement of MEC applications can give service providers a competitive advantage, but diverse industries have different requirements, including security and privacy concerns. Security SLAs for MEC hosting should isolate sensitive workloads, as malicious applications can be deployed among benign ones. Privacy issues arise, such as in MEC applications involving VR/XR, where users' environmental information could reveal their whereabouts.

\noindent\textbf{Regulatory privacy non-compliance of constrained data movement.} 
The EU's General Data Protection Regulation (GDPR) {\color{black}\citep{hoofnagle2019european}} is in place to regulate data leakage and transfer to third parties, including cloud providers. Compliance with local laws for roaming user data is crucial, with non-compliance recorded for liability purposes. Virtual On-Board Units (vOBUs) enforce GDPR compliance and must be adaptable to different legal contexts while protecting the user equipment-to-cloud channel. Their actions must be stored securely in the operator's infrastructure.

\noindent\textbf{Falsified command signaling attack to drones/robots/vehicles.} 
Cooperative, connected, and automated mobility (CCAM) services for 5G/6G vehicular environments face security risks from remote attacks that manipulate vehicle control. This threat concerns teleoperated driving scenarios, where vehicles receive driving commands over mobile networks. Such attacks could be used to manipulate safety-critical vehicle components and falsify sensor data. Strong security measures, e.g., vulnerability patching and secure communication protocols, must be implemented.

\noindent\textbf{Falsified information broadcast in 6G.} 
The possibility of falsified information being broadcast in 5G/6G networks is a serious threat to many applications, including autonomous driving. HD maps, which are crucial for safe and efficient autonomous driving, could be manipulated or falsified by attackers, posing risks to the safety and security of both the driver and vehicle. This scenario relies on cooperation between autonomous vehicles and communication network infrastructure, making information exchange crucial. It also opens the door to unauthorized tracking of vehicles.

\noindent\textbf{Security and privacy concerns in personal IoT network (PIN).} 
6G networks may be exposed to data breaches, malware, eavesdropping, and physical attacks. Personal IoT (PIoT) networks such as PINs and CPNs could be targeted by attackers to gain unauthorized access to sensitive data or disrupt system operations. The collection of personal data from PIoT networks also raises privacy concerns. To mitigate these risks, strong security measures such as encryption and access controls, as well as transparent privacy protections, will need to be implemented.

\noindent\textbf{Privacy risks of 6G as a sensor.} 
The ``Network as a Sensor" uses communication signals to sense the environment by reflecting signals off objects and surfaces. This technology 
may raise privacy concerns if it collects information on individuals. 
Measures need to be taken to regulate data collection and usage.
Empowered by ML, 6G systems can be vulnerable to privacy-related attacks if not designed and implemented with privacy in mind \citep{0009}. 
Table \ref{tab: risk assessment for novel protocols} maps the potential attacks, vulnerabilities, and risks onto the 6G new protocols and application scenarios. 

\begin{figure*}
\small
\begin{longtable}{|l|l|l|l|l|l|l|l|}
\caption{Risk of the attacks for novel 6G protocols and application scenarios.}
\label{tab: risk assessment for novel protocols}			
\\			\hline 
		& \multicolumn{1}{l|}{\textbf{NAD-FL}}               
            & \multicolumn{1}{l|}{\textbf{SP-MEC}}               
            & \multicolumn{1}{l|}{\textbf{\begin{tabular}[c]{@{}l@{}} Privacy \\ Compl.\end{tabular}}}                
            & \multicolumn{1}{l|}{\textbf{\begin{tabular}[c]{@{}l@{}} Falsified \\ command\end{tabular}}}                      
            & \multicolumn{1}{l|}{\textbf{\begin{tabular}[c]{@{}l@{}}Falsified \\ broadcast\end{tabular}}} 
            & \multicolumn{1}{l|}{\textbf{\begin{tabular}[c]{@{}l@{}}PIN\end{tabular}}} 
            & \multicolumn{1}{l|}{\textbf{\begin{tabular}[c]{@{}l@{}}Net\\ Sen.\end{tabular}}}   \\           
			\hline 
			\endfirsthead
\begin{tabular}[l]{@{}l@{}}Background  Knowledge  Attacks \end{tabular}    &  &    \multicolumn{1}{c|}{\checkmark}   &   &   &  \multicolumn{1}{c|}{\checkmark}  &  \multicolumn{1}{c|}{\checkmark}  &      \\ \hline
\begin{tabular}[l]{@{}l@{}}Inference  Attacks\end{tabular}   & & \multicolumn{1}{c|}{\checkmark}   &  &  & &  \multicolumn{1}{c|}{\checkmark}  & \multicolumn{1}{c|}{\checkmark}          \\ \hline

\begin{tabular}[l]{@{}l@{}}Linkage  Attacks\end{tabular}   &  & \multicolumn{1}{c|}{\checkmark}   &  &   &   & \multicolumn{1}{c|}{\checkmark}    &   \\ \hline
\begin{tabular}[l]{@{}l@{}}Privacy  leakage  in digital  representations\end{tabular} &  &\multicolumn{1}{c|}{\checkmark}    & \multicolumn{1}{c|}{\checkmark}   &  & \multicolumn{1}{c|}{\checkmark}    & \multicolumn{1}{c|}{\checkmark}   & \\ \hline
\begin{tabular}[l]{@{}l@{}}Privacy risks  of integrated  sensing and  \\communication\end{tabular}      &  &  &  & \multicolumn{1}{c|}{\checkmark}   & \multicolumn{1}{c|}{\checkmark}   &  & \multicolumn{1}{c|}{\checkmark}   \\ \hline
\begin{tabular}[l]{@{}l@{}}Privacy  leakage in computation  offloading\end{tabular} &  & \multicolumn{1}{c|}{\checkmark}   & \multicolumn{1}{c|}{\checkmark}   &  &   & \multicolumn{1}{c|}{\checkmark}   &  \\ \hline
\begin{tabular}[c]{@{}l@{}}Privacy risks in distributed  learning\end{tabular} & \multicolumn{1}{c|}{\checkmark}   & \multicolumn{1}{c|}{\checkmark}   & \multicolumn{1}{c|}{\checkmark}   &  &  & \multicolumn{1}{c|}{\checkmark}   &  \\ \hline
\end{longtable}
 \footnotesize{NAD: Network anomaly detection, SP: secure and private}
 
\end{figure*}



\subsection{Privacy and security features inherited from 5G}
5G has already implemented advanced privacy and security protection mechanisms. 
In this section, we delve into the existing methods for risk mitigation
and highlight key security and privacy features inherited from 5G. 

\subsubsection{Defeating DDoS Attacks by Managing Shared Resources in an 
Intelligent and Secure Way} 

Distributed Denial of Service (DDoS) attacks are becoming increasingly covert, capable of imitating legitimate behavior while using low bandwidth. This makes them harder
to detect. Additionally, the interconnected nature of virtual network functions
and shared infrastructure resources in 5G networks increases the risk of indirect
DDoS attacks, where the exhaustion of resources in one slice can impact the 
availability and performance of services in other slices. To combat this, the 
platform must have fail-safe mechanisms in place to protect shared resources 
from depletion. For instance, if the resources of Slice B are allowed to scale
up or out without proper controls, it could lead to the depletion of shared 
resources, resulting in decreased performance for Slice A or hindering the 
ability to create a new Slice C.

The target is to address instances where undetected attacks on a slice lead 
to resource depletion in shared infrastructure, impacting other critical 
slices. Although it doesn't directly neutralize the threat, it helps safeguard shared resources and reduce the impact on unaffected slices or services. 
The roles involved include: (1) Malicious party (Mallory), (2) Mobile Network 
Operator (MNO): RAN, 5G Core (CP + UP), Mobile Edge, (3) Legitimate mobile 
device users (Alpha, Bravo), and (4) Malicious mobile users (Yankee, Zulu). 
The basic flow of this use case consists of the following steps. Imagine two services, A and B, operating within a 5G core along the user's data path. The resources dedicated to these services are confined within two distinct slices that span from the devices through the RAN domain to the Core Domain of the MNO infrastructure. These services are governed by specific Service Level Agreements (SLAs).

\begin{figure*}[t]
    \centering
    \includegraphics[width=0.6\textwidth]{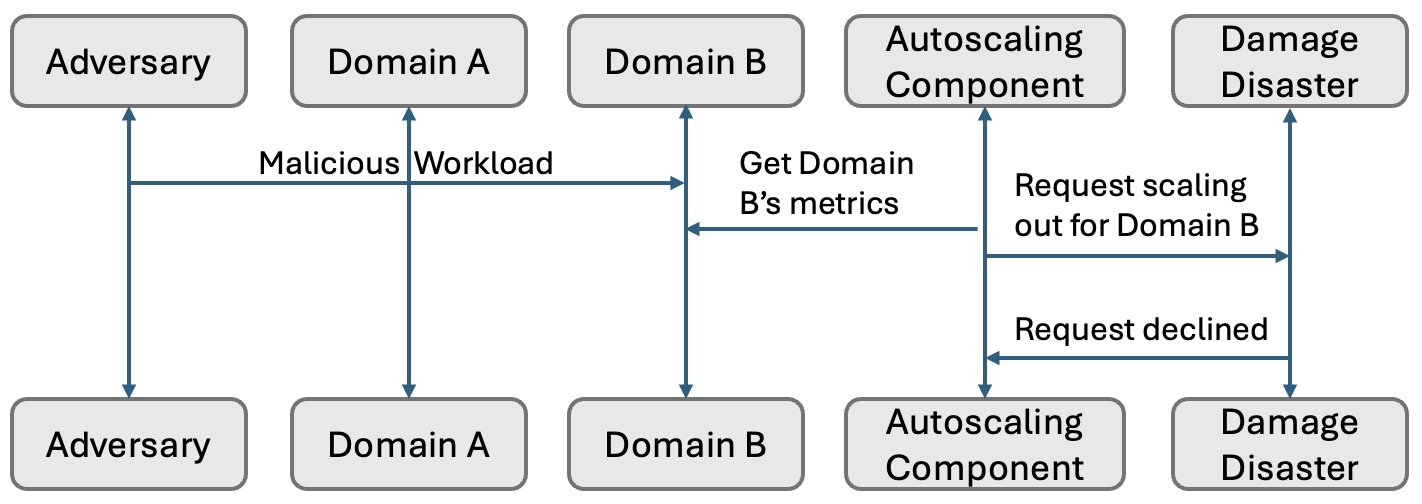}
    \caption{Diagram and workflow for defeating DDoS attacks. }
    \label{fig:threat_1}
\end{figure*}

The malicious actor, Mallory, launches a stealthy DDoS attack on service B using compromised devices connected to slice B. The existing security measures (such as a firewall or intrusion detection system) are unable to quickly differentiate between malicious and legitimate traffic. The attack affects the SLAs of slice B, leading the system to 
repeatedly engage in auto-scaling operations, such as increasing resources or the 
number of virtual machines serving the service. These repeated auto-scaling 
operations can exhaust resources shared with slice A, such as CPU, memory,
network queues, application caches, disk I/O, and file descriptors. For instance, 
the resource pools managed by the RAN may be exhausted in favor of the 
malicious slice. A damage control mechanism should mitigate the impact on slice B by verifying and potentially preventing new resource allocations, as illustrated in Fig.~\ref{fig:threat_1}.

\subsubsection{Division of Core Sectors of Virtualised Environments}

This use case examines the application of Trust and Liability management principles within a virtualized infrastructure in the context of a 5G ecosystem. Legal and regulatory frameworks, along with industry-specific requirements, highlight the need for 5G and B5G infrastructures to meet various obligations (such as those from the Cybersecurity Act {\color{black}\citep{CybersecurityAct}}, NIS directive {\color{black}\citep{nis2d}}, and standards relevant to 5G verticals like eHealth, Transport, Energy, Vehicular sectors, and those governed by the Seveso directive {\color{black}\citep{SevesoDirective}}), while also being adaptable. Implementing the highest security level, as outlined in the Cybersecurity Act (which defines three assurance levels: Basic, Substantial, and High), is impractical. Some requirements may conflict, and many use cases do not require the maximum security level. Verticals are unlikely to pay for services they neither need nor use. Ensuring such a high-security level across all components of the network and services can be prohibitively expensive, potentially raising configuration costs to unsustainable levels.

Currently, in virtualized environments, there is a lack of proper isolation for
critical services from basic services. This leads to the need for dedicated
physical infrastructure to run critical services. By formalizing security 
requirements through colocalization constraints or criticality levels, we can 
use placement optimization algorithms to orchestrate physical 
resources to meet the needs of both critical and basic services. This algorithm
considers various constraints and has the potential to optimize for factors, 
such as latency and energy consumption across multi-site infrastructures in the
future.
To ensure proper isolation of critical services from basic services,
it is important to demonstrate that this isolation has been 
achieved without giving direct physical access to the infrastructure. The Deep Attestation framework  {\color{black}\citep{8980316}} provides a solution by allowing the Client to agree
on a method for measuring this isolation. 
Services are isolated from each other based on their affinity or criticality
constraints. 
The Deep Attestation framework then uses this information and attestation scheme to digitally sign 
and protect the security properties on each targeted server. 

Several roles are involved in this process: (1) the Virtualized Infrastructure Operator, (2) the Client or Vertical that requires isolation for its critical services, (3) a deep attestation framework that generates evidence, and (4) a placement optimization system integrated with the infrastructure orchestration system. 
When starting the process, the Infrastructure Operator arranges the components based 
on isolation requests and uses placement optimization under security constraints.
 The isolation is assessed using a pre-determined measurement method, and the Client receives proof of the isolation. During the operational phase, the Client may request evidence of isolation. Each request prompts a measurement of the isolation using the agreed method. The corresponding proof is provided to the Client. 

\subsubsection{Security Performance Evaluation}
In this scenario, a vertical user relies on an infrastructure supplied by an Infrastructure Operator to run their service. To ensure the service runs 
perfectly and securely, the vertical may require certain properties from the 
infrastructure, specifically security properties such as installing an 
antivirus on a specific node or having a specific boot configuration. 
These properties are crucial for the service to run properly, so verifying and monitoring them is important. One way to do this is by directly 
challenging the infrastructure, but this can result in security issues 
such as leaks of private information or an increase in attack vectors.

To mitigate the previously mentioned security concerns, the vertical can assign the responsibility of managing monitoring software to the infrastructure operator. This is achieved through an agreement that outlines the necessary infrastructure properties and the methods for data collection. The infrastructure operator can then demonstrate that the appropriate software has been utilized for data collection and provide proof through a Remote Attestation (RA) mechanism  {\color{black}\citep{kylanpaa2016remote}}. 
The purpose of this use case is to ensure that the infrastructure adheres to the security properties agreed upon by both the vertical and the infrastructure operator while preserving the privacy and control of the infrastructure and its owner.  The roles include: (1) vertical user, (2)
the RA server, and (3) RA agents. In the process, the vertical user will utilize the infrastructure to deliver a service.

The RA server oversees the attestation service and is the sole point of access for obtaining information or evidence regarding the infrastructure. It has the ability to (a) assess whether a target, such as a node within the infrastructure, is capable
of performing an attestation, (b) install the necessary software on 
the target and set it up, (c) remove RA software from a target, 
(d) provide a list of active RA targets, and (e) execute an attestation on a target and validate the results. The RA server manages RA Agents 
to carry out these actions. RA agents collect data from the infrastructure
and securely compute and deliver attestation results. 
They reside within the infrastructure domain at the virtualized layer or the virtualization management layer, with the ability to access the hardware, including 
the root of trust.

\subsubsection{Attacks on Encrypted Network Traffic and Defense}

Network operators rely on network traffic monitoring to improve 
the efficiency of their platforms. However, with the trend of encryption 
in internet traffic, existing management and security mechanisms are 
impacted. In 5G network, data plane traffic between the RAN and Core relies on GTP-U which is not always encrypted, but IPsec is being adopted to 
increase security. 5G Core also includes the Service Based Architecture (SBA), 
which uses HTTP/2 {\color{black}\citep{pollard2019http}} as the protocol base for signalling traffic instead of 
the legacy DIAMETER protocol. 3GPP mandates TLSv1.2 for RESTful APIs in 
cloud environments and microservices, making traditional network security tools, such as deep packet 
inspection (DPI) {\color{black}\citep{bremler2014deep}}, less effective in detecting common attacks. 

Cybercriminals are increasingly using encryption in their communication and attack methods, complicating the detection of malicious activities. This shift brings about new risks, particularly through REST API channels concealed within TLS, including malware actions, DDoS attacks, and assaults on SBA microservices. Furthermore, the implementation of 5G Core elements and monitoring systems using microservices, NFV, and cloud-based approaches can leave them 
vulnerable to introspection attacks which can disable detection.
%
Placing multiple probes in key areas of the 5G network, whether through virtual or physical taps, can trigger security alerts that orchestration systems can monitor and respond to. Moreover, it involves employing data and software protection methods, such as Trusted Execution Environment (TEE) technologies like Intel's SGX enclave, to block unauthorized access and detect software behavior.

In this use case, the following roles are involved. (1) 5G Network Administrator, 
who is responsible for managing and securing the network, including monitoring
and remediation. They are able to enforce specific policies in the network. 
(2) Malicious party who aims to compromise the system and steal or alter 
information. They may try to hide their activity in encrypted traffic or 
attack the components. (3) 
Security monitoring probes, designed to be compatible with the INSPIRE-5Gplus (Infrastructure for Spatial Information in the European Community) framework {\color{black}\citep{inspire5gplus2023}}, gather crucial information and transmit it to analytics engines that identify potential attacks. (4) Security Management
Domain, which provides security analytics, decision-making, visualization, 
reporting of attacks, and enforcement of corrective actions.

In this attack, the attacker gains control of some network resources. The network 
administrator notices a decline in service performance due to the attacker's 
actions but is uncertain of the cause or solution. The administrator deploys
AI-based monitoring agents (Smart Traffic Analyzer) in the network to monitor
encrypted control and data planes. To protect against unauthorized access, 
the administrator increases security on the network probes in untrusted 
environments such as third-party data centers. To prevent introspection attacks,
the administrator uses TEE to verify the 
integrity of the network functions.
The deployed security agents (probes) gather metrics and utilize AI/ML algorithms to detect malicious behaviour patterns within encrypted traffic. Once malicious activities are detected, they are reported to the administrator, who can then respond by implementing security measures, such as configuring firewalls or deploying active probes. Additionally, compromised functions can be sanitized and re-deployed to eliminate the threat.

\subsubsection{Secure Network Analytics Data Usage}

To detect mobile device-related threats, 5G networks use active or passive 
probes 
that collect data and send it to security analytics functions for analysis. 
These functions, such as the Network Data Analytics Function (LATERDAF), can 
identify malicious behavior and are part of the Service Based Architecture 
in 3GPP TS 23.501. These analytics can be used within the 
framework to detect and manage current or potential attacks at both the 
domain and end-to-end security management levels.

The goal of this use case is to showcase how cybersecurity analytics, generated by specialized 5G core network functions, can be integrated into the security architecture. It demonstrates the advantages that external management systems can gain from the data provided by 5G systems. The key roles include: (1) MNOs who manage and operate a 5G network equipped with a 5G Core and LATERDAF, (2) Mobile device users (e.g., Alice and Bob) who connect to the MNO network using 5G devices, (3) A malicious actor (Mallory) who attempts to carry out cyber or physical attacks on mobile devices, such as hijacking or theft, and (4) The security management system for the UE domain.

\subsubsection{Leveraging Similarity Learning to Conduct Root Cause Analysis }

In 5G-based systems, recurring failures are a frequent occurrence. Although experienced system administrators can typically resolve these issues swiftly and effectively, managing such failures becomes increasingly complex and time-consuming in more intricate systems. This complexity arises because failures tend to propagate through causal chains, leading to evolving symptoms. To address this challenge, an automated tool is needed to aid in troubleshooting by grouping causally related events and distinguishing them from unrelated ones. However, this can be difficult, as different system components may exhibit similar symptoms for failures that are not related.

An automated Root-Cause Analysis (RCA) tool is essential for effectively troubleshooting complex systems, such as those in Information and Communication Systems (ICS) and 5G-based networks. RCA examines highly detailed monitoring indicators, including statistics and data extracted from logs and network traffic, to assess the similarity of new events to previous incidents. By leveraging past experiences and building a historical database, RCA facilitates timely and effective remediation actions to prevent or mitigate recurring issues. In an Industrial Campus IoT network, an anomaly detection system, combined with an ML-based RCA, will be deployed to analyze traffic and various hardware indicators, pinpointing the causes of anomalies and distinguishing between false positives and true negatives.

\begin{figure*}[!t]
    \centering
    \includegraphics[width=0.6\textwidth]{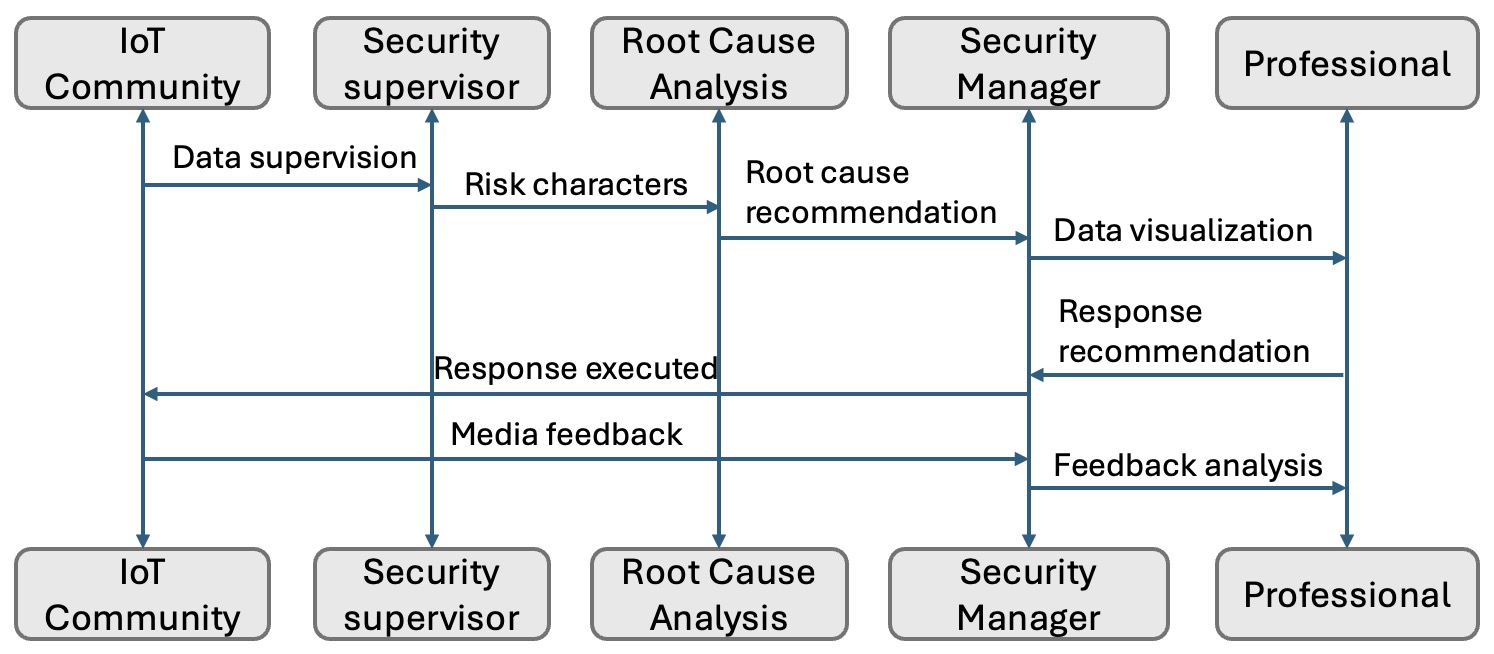}
    \caption{Diagram and workflow of root cause analysis.}
    \label{fig:threat_6}
\end{figure*}

The process in this use case unfolds as follows: The IoT Campus sends monitoring data, including information from IP cameras, to the Multi-Modal Traffic Root Cause Analysis (MMT-RCA) module for analysis. The MMT monitoring framework, encompassing the MMT-RCA module, processes the collected data and raises alerts when necessary. The Security Orchestration system then receives the analysis results from the MMT-RCA module and presents them on web-based dashboards. Upon receiving an alert from the MMT-RCA module, the Technician Command Center can activate CRITICAL mode by pressing a button on the web interface. Fig.~\ref{fig:threat_6} depicts the steps involved in this use case.

\subsection{Risk assessment frameworks for 6G}
\label{sec:risk-assessment-6g}

In this section, we provide an overview survey of existing frameworks proposed for 
earlier communication networks, such as 5G and 4G, especially with a focus on 
determining how frameworks differentiate between risk in the networks themselves, 
the observability of the networks, and the application areas enabled by the network
technology. 
We identify potential limitations or challenges of 
relying solely on these existing risk assessment frameworks in 6G.

\subsubsection{EU Cybersecurity Risk Assessment and Toolbox on 5G Networks}

The EU Toolbox for 5G Security {\color{black}\citep{eutoolbox5gsecurity}} outlines a set of comprehensive measures aimed at ensuring a coordinated approach to 5G network security across the EU~\citep{eu-risk-framework}. It addresses cybersecurity risks identified in the EU risk assessment report. Based on EU Commission Recommendation 2019/534, it emphasizes three technical aspects: 1) The shift towards software-based solutions and virtualization, utilizing technologies such as Software Defined Networking (SDN) and Network Functions Virtualisation (NFV); 2) The implementation of Network Slicing; and 3) Enhanced functionality at the network edge with a move towards a more decentralized architecture. 
The framework follows ISO/IEC 27005, 2018 risk assessment methodology, assessing threats to confidentiality, availability, and integrity, such as disruptions to 5G networks, traffic manipulation, and the risk of cascading impacts on other infrastructure systems vis 5G.

The 5G Toolbox is a key component of 5G security~\citep{friis2021}. It encompasses strategic measures (SM), technical measures (TM), and supporting actions. 
Strategic measures target local regulators, often requiring national laws for effectiveness, while technical measures are implemented by mobile network operators and partners like cloud providers and service vendors. Supporting actions help meet security goals, with risk mitigation strategies tailored to the risk level, which may be classified as very high, high, medium, or low. However, the Toolbox’s application varies across EU member states and does not consider business models like Infrastructure as a Service (IaaS) or Software as a Service (SaaS).

The toolbox may not fully address 6G-based scenarios, as it overlooks risks emerging from enhanced 6G capabilities like machine-to-machine communication, advanced sensing, and AI-driven connectivity. For instance, 6G aims to realize the Internet of Everything (IoE), which will involve billions of interconnected and diverse devices. The hyper-connected nature of IoE in 6G applications could potentially lead to significant privacy concerns. Data theft exploiting resource-limited IoT devices may compromise data privacy, location privacy, and identity privacy. Consequently, the toolbox may overlook critical factors that contribute to high-risk scenarios within 6G domain applications.

\subsubsection{5G-ENSURE (5G PPP)}

5G-ENSURE is a project initiated by 5G PPP to explore risk assessment and mitigation strategies for 5G networks~\citep{arfaoui2017}. The risk assessment methodology employed by 5G-ENSURE is grounded in the Risk Management Process outlined in ISO 27005, with a particular focus on the simplified version found in NIST SP-800-30~\citep{NIST-SP}. The project introduced a conceptual security framework for 5G, incorporating various use cases that highlight the security and privacy challenges within 5G networks~\citep{5g-ensure-usecases}. 

The 5G-ENSURE risk assessment process, adhering to the ISO 27005 standard, consists of three key stages: (1) Identifying risks, (2) Analyzing risks, and (3) Evaluating risks. Risk identification seeks to pinpoint potential factors that could lead to loss or damage, as well as the circumstances under which such events might occur. The framework initially categorizes the assets within a 5G mobile network, followed by an examination of the threats that could jeopardize these assets. It then estimates the potential impact if any of these threats were to materialize.

However, incorporating 5G-specific assets and vulnerabilities into this framework poses challenges, as each identified risk scenario depends heavily on accurately assessing both the likelihood of occurrence and the potential impact, which requires domain-specific expertise. To address this, the authors proposed three main approaches. 
The first involves adapting evaluations from 4G systems, though these are often inadequate for 5G's advanced capabilities. The second uses theoretical values from existing literature, but these are similarly based on 4G methods, raising doubts about their relevance to 5G. The third relies on expert estimates from 5G PPP projects, though mobile network operators may still need to fine-tune these estimates to fit real-world 5G scenarios.

\subsubsection{Interoperable EU Risk Management Toolbox}

The EU risk management (RM) toolbox, introduced by ENISA~\citep{iermt}, addresses concerns about interoperability in the use of information security risk management methods. This toolbox offers a reference framework designed to facilitate the interpretation, comparison, and aggregation of outcomes from various risk assessment approaches. It enables different stakeholders to collaborate on common threats and risk scenarios, and to compare risk levels, even when these risks are assessed using different or proprietary tools and methodologies. 

Developed under the ENISA Interoperable EU Risk Management Framework~\citep{enisa-eu-framework}, which was published in December 2022, the toolbox is grounded in the basic risk management processes defined in ISO/IEC 27005:2018 and the EU Information Technology Security Risk Management Methodology (ITSRM2)~\citep{enisa-itsrm}. 

Key components of the toolbox include the identification and categorization of assets, threat identification, the description of attack scenarios, and the assessment and comparison of risk levels. It also provides a common vocabulary to ensure that stakeholders can understand and compare the outcomes of different RM methods. This enables organizations to align their results with those of others, even when different RM tools are used.

It should be noted that the majority of threat scenarios in this toolbox focus on cybersecurity, with limited attention to user and data privacy risks. 
Although the toolbox provides classifications of assets and threats that can be mapped to various risk scenarios within an organization, it would benefit from an iterative evolution process to encompass the broader spectrum of risks associated with 6G applications. Consequently, a common set of measures must be developed and aligned with different risk management methods to ensure consistent and unified risk treatment. 
Additionally, 
the current methodology is asset-focused, lacking support for scenario-based assessments, and comparing risk frameworks directly may lead to inaccurate conclusions since organizations' risk scores may not be directly comparable.

\subsubsection{ENISA personal data processing operation tool}

In its 2017 guidelines~\citep{aar2017}, ENISA introduced a streamlined approach designed to assist small to medium enterprises (SMEs), whether they act as data controllers or processors, in navigating their specific data processing operations. This guidance helps them assess relevant security risks and adopt appropriate security measures accordingly. Rather than introducing a new risk assessment methodology, the personal data risk assessment tool builds on established practices in the field, offering tailored support for SMEs. 
When it comes to managing information security risks versus personal data security risks, the evaluation of risk levels is conducted from two distinct perspectives: the first emphasizes the impact on the data controller, while the latter focuses on the impact on data subjects. The toolbox's risk evaluation primarily considers the effects on the data controller, often overlooking the potential impact on data subjects unless that impact also directly affects the data controller.

Currently, the toolbox offers a set of web-based questionnaires for users to assess risks within their data processing environments. However, these forms are static and do not allow users to tailor questions to their specific use cases or application scenarios. This limitation reduces the tool’s adaptability and practical application in diverse real-world situations. Additionally, for less experienced users, such as those in non-technical fields, the questions may not provide sufficient context or clarity to enable a precise and meaningful evaluation. For instance, a question like "Can unauthorized individuals easily access the data processing environment?" might be challenging for a user to answer accurately without an understanding of potential vulnerabilities or attacks relevant to their environment.

Moreover, the impact assessment component of the tool asks users to answer three questions regarding the potential impact on confidentiality, integrity, and availability, with options to rate the impact as Low, Medium, High, or Very High. However, these questions do not account for critical factors such as the type of personal data involved, the significance of the processing operation, the volume of data, specific characteristics of the data controller, or the categories of data subjects, all of which could significantly influence the level of impact. As a result, the tool may fail to capture the nuanced risk levels associated with different contexts and scenarios.

\subsubsection{SERIMA}

The Luxembourg Institute of Science and Technology (LIST) has developed a prototype known as SERIMA (Security Risk Management)~\citep{serima}, a regulatory platform aimed at helping telecommunications operators conduct risk analyses, particularly within the telecommunications industry. This platform is designed to provide operators with a standardized methodology for performing risk assessments and reporting incidents in compliance with current regulations. SERIMA addresses challenges such as discrepancies in format, methodology, data reconciliation, and the comparison and analysis of data.

SERIMA is an enhanced version of the TISRIM tool~\citep{TISRIM}, which was initially created by LIST at the request of the Luxembourg Regulatory Institute (ILR). In its first phase, the platform will be utilized exclusively by operators in the electronic communications sector. However, in alignment with the EU NIS directive, its application will gradually expand to other sectors, including energy, transportation, healthcare, digital infrastructure, and the supply and distribution of drinking water. To facilitate this expansion, ILR's Network and Information Systems Security (NISS) department has established working groups to customize the platform for each specific sector.

\subsubsection{Open Radio Access Network (O-RAN) Risk Analysis (5GRANR)}

Commissioned and financed by the Federal Office for Information 
Security (BSI)  in Germany, a group of researchers from the Barkhausen 
Institute  in Germany conducted a study on analyzing the security 
risks that arise from the O-RAN implementation of a 3GPP RAN as 
specified by the O-RAN Alliance. This study~\citep{oran} presents a risk 
analysis considering confidentiality, integrity, accountability, 
availability, and privacy on O-RAN. 
The risk analysis essentially only considered threats and 
vulnerabilities that are specific to 3GPP or O-RAN. The 
considerations made are thus limited to the RAN interfaces to the 
user equipment and the 5G core.  
The study examined risks from three perspectives: (1) End users, focusing on risks to user data and network usage; (2) 5G network operators, evaluating threats to network operations and assets; (3) Society/government, considering the impact on critical services like electricity, healthcare, and transport.


The three stakeholder perspectives analyze 5G RAN risks from different angles, but the analysis is based on 3GPP and O-RAN Alliance standards, not on a specific 5G RAN implementation. These standards provide a framework within which compliant 5G O-RAN implementations can operate.

In the United States, similar to initiatives in the EU, there are several efforts and reports focused on addressing 5G-related risks and implementing corresponding security measures. The Cybersecurity and Infrastructure Security Agency (CISA), in collaboration with the National Security Agency (NSA) and the Office of the Director of National Intelligence, as part of the Enduring Security Framework (ESF), published the ``Potential Threat Vectors to 5G Infrastructure''~\citep{cisa-threat-vector} report, which assesses 5G-related risks. Additionally, other key documents, such as the National Strategy to Secure 5G and its accompanying Implementation Plan {\color{black}\citep{nationalstrategy5g}}, delve into risks related to the supply chain and the impact of standardization on 5G security.

In a separate study, Laaroussi and colleagues~\citep{laaroussi2022} from the Ericsson Research Centre in Finland conducted a comprehensive risk assessment focused on telepresence holographic communication within a 6G environment. Their analysis utilized the STRIDE methodology {\color{black}\citep{khan2017stride}}, categorizing threats into Spoofing, Tampering, Repudiation, Information Disclosure, Denial of Service (DoS), and Elevation of Privilege. These threats were assessed across three attacker profiles: Insiders, Hackers, and Outsiders. Similarly, researchers from the University of Alabama performed a study on video conferencing security~\citep{hasan2021}, identifying potential attackers as end users or administrators motivated by factors such as financial gain, espionage, business competition, or personal ideology. The study highlighted threats including content spoofing, tampering with storage or logs, data breaches, DoS attacks, and unauthorized access.

\subsection{Discussion and Findings}

Data privacy has been a concern for decades, with regulations like GDPR and the EU Right to be Forgotten ensuring protection, the EU Right to be Forgotten, and the GDPR, mandating the safeguarding of personal information. However, as we move towards 6G, it is expected that key applications, including wearable devices, will increasingly handle highly sensitive user data \citep{batalla2020, sirohi2023}. In densely connected networks, advancements in localization technology could enable highly accurate tracking of a user’s movements, raising concerns that personal data might be exploited for surveillance purposes. Moreover, the widespread adoption of cloud-based infrastructure and applications increases the risk of unauthorized access and potential exposure of sensitive data. Additionally, protecting personal data in an era dominated by AI will present new challenges, as AI-based techniques could be used to target sensitive information.

In 6G networks, protecting data related to user identity, behavior, and user-generated content is critical for privacy. Only authorized entities should have the ability to interpret information that reveals a user’s identity or behavioral patterns. Unique aspects of user identity and behavior in 6G are tied to a unified identity definition and the structure of signaling messages. While user-generated data is not stored within the telecom network itself, it must be secured during data processing and operations using methods such as encryption and robust security management practices.

\begin{table*}[t]
\centering
\caption{Summary of functionalities provided by different risk frameworks.}
\label{tab:risk-framework-summary}
\begin{small}
\begin{tabular}{|l|ccccccc|}
\hline
\multirow{2}{*}{Capability} &
  \multicolumn{7}{c|}{Risk framework or toolbox} \\ \cline{2-8} 
 &
  \multicolumn{1}{c|}{EUT5G} &
  \multicolumn{1}{c|}{\begin{tabular}[c]{@{}c@{}}5G \\ ENSURE\end{tabular}} &
  \multicolumn{1}{c|}{\begin{tabular}[c]{@{}c@{}}EU Risk \\ Management\\ Toolbox\end{tabular}} &
  \multicolumn{1}{c|}{\begin{tabular}[c]{@{}c@{}}ENISA\\ Personal\\ data processing\\ operation tool\end{tabular}} &
  \multicolumn{1}{c|}{SERIMA} &
  \multicolumn{1}{c|}{\begin{tabular}[c]{@{}c@{}}O-RAN\\ Risk analysis\end{tabular}} &
  \begin{tabular}[c]{@{}c@{}}CISA \\ Enduring\\ Security \\ Framework\end{tabular} \\ \hline
\begin{tabular}[c]{@{}l@{}}Considered \\ network security\end{tabular} &
  \multicolumn{1}{c|}{\checkmark} &
  \multicolumn{1}{c|}{\checkmark} &
  \multicolumn{1}{c|}{\checkmark} &
  \multicolumn{1}{c|}{} &
  \multicolumn{1}{c|}{\checkmark} &
  \multicolumn{1}{c|}{\checkmark} & \checkmark
   \\ \hline
\begin{tabular}[c]{@{}l@{}}Consider application \\ security\end{tabular} &
  \multicolumn{1}{c|}{\checkmark} &
  \multicolumn{1}{c|}{\checkmark} &
  \multicolumn{1}{c|}{} &
  \multicolumn{1}{c|}{\checkmark} &
  \multicolumn{1}{c|}{\checkmark} &
  \multicolumn{1}{c|}{} & \checkmark
   \\ \hline
\begin{tabular}[c]{@{}l@{}}Consider data privacy\\ in the network\end{tabular} &
  \multicolumn{1}{c|}{\checkmark} &
  \multicolumn{1}{c|}{\checkmark} &
  \multicolumn{1}{c|}{\checkmark} &
  \multicolumn{1}{c|}{} &
  \multicolumn{1}{c|}{} &
  \multicolumn{1}{c|}{\checkmark} &
   \\ \hline
\begin{tabular}[c]{@{}l@{}}Consider data privacy in\\ user applications\end{tabular} &
  \multicolumn{1}{c|}{} &
  \multicolumn{1}{c|}{} &
  \multicolumn{1}{c|}{} &
  \multicolumn{1}{c|}{\checkmark} &
  \multicolumn{1}{c|}{} &
  \multicolumn{1}{c|}{} &
   \\ \hline
\begin{tabular}[c]{@{}l@{}}Consider different \\ threat models or threat \\ vectors\end{tabular} &
  \multicolumn{1}{c|}{} &
  \multicolumn{1}{c|}{\checkmark} &
  \multicolumn{1}{c|}{\checkmark} &
  \multicolumn{1}{c|}{\checkmark} &
  \multicolumn{1}{c|}{\checkmark} &
  \multicolumn{1}{c|}{\checkmark} & \checkmark
   \\ \hline
\begin{tabular}[c]{@{}l@{}}Applicable to different\\ use cases\end{tabular} &
  \multicolumn{1}{c|}{} &
  \multicolumn{1}{c|}{\checkmark} &
  \multicolumn{1}{c|}{\checkmark} &
  \multicolumn{1}{c|}{} &
  \multicolumn{1}{c|}{} &
  \multicolumn{1}{c|}{\checkmark} &
   \\ \hline
\begin{tabular}[c]{@{}l@{}}Can be adopted to\\ different risk scenarios\end{tabular} &
  \multicolumn{1}{c|}{\checkmark} &
  \multicolumn{1}{c|}{\checkmark} &
  \multicolumn{1}{c|}{} &
  \multicolumn{1}{c|}{} &
  \multicolumn{1}{c|}{} &
  \multicolumn{1}{c|}{\checkmark} &
   \\ \hline
\begin{tabular}[c]{@{}l@{}}Require domain\\  knowledge\end{tabular} &
  \multicolumn{1}{c|}{\checkmark} &
  \multicolumn{1}{c|}{\checkmark} &
  \multicolumn{1}{c|}{\checkmark} &
  \multicolumn{1}{c|}{\checkmark} &
  \multicolumn{1}{c|}{\checkmark} &
  \multicolumn{1}{c|}{} & \checkmark
   \\ \hline
\begin{tabular}[c]{@{}l@{}}Applicability to 6G\\ environments (based on the \\ different use cases and \\ technologies \\ that are possible in a \\ 6G environment)\end{tabular} &
  \multicolumn{1}{c|}{} &
  \multicolumn{1}{c|}{Limited} &
  \multicolumn{1}{c|}{Limited} &
  \multicolumn{1}{c|}{} &
  \multicolumn{1}{c|}{} &
  \multicolumn{1}{c|}{Unsure} &
   \\ \hline
\end{tabular}%
\end{small}
\end{table*}


The risk assessment framework for 6G must be carefully developed by reviewing current methodologies and adapting or creating a comprehensive model suited to the new network architecture, stakeholders, and business models.  As outlined in Table~\ref{tab:risk-framework-summary}, existing frameworks do not fully address 6G capabilities, necessitating a new approach that thoroughly assesses potential privacy leaks and vulnerabilities in 6G data management. Prior to sharing data with third parties or introducing new data types, careful consideration of user privacy is essential. Additionally, new mechanisms may be needed to address potential privacy threats from unseen connections among existing data.

\section{International perspectives on 6G technology} \label{ipt}
\subsection{Industrial vision and activities}

\subsubsection{Industrial visions regarding 6G}

6G has been envisioned as a key enabler of many new and innovative applications and services across multiple industries, 
driving economic growth and improving quality of life. We reviewed several world-learning industrial players in 6G, including Nokia, Ericsson, DOCOMO, Samsung, InterDigital and Qualcomm. Though they differ in focus areas, most companies prioritize ultra-high-speed, low-latency communication, massive machine-type communications, and AI alongside ML. Security and privacy, as well as energy efficiency, also emerge as key considerations for many companies. 
Table~\ref{tab:fa_industry} summarizes their focal areas.

    
\begin{small} 
\begin{longtable}{|l|l|l|l|l|l|l|}	
\caption{Focus areas - 6G industry}
\label{tab:fa_industry}

\\		\hline 

		\multicolumn{1}{|l|}{\textbf{Focus Areas}}               
            & \multicolumn{1}{l|}{\textbf{Nokia}}               
            & \multicolumn{1}{l|}{\textbf{Ericsson}}    
            & \multicolumn{1}{l|}{\textbf{DOCOMO}}               
            & \multicolumn{1}{l|}{\textbf{Samsung}}  
            & \multicolumn{1}{l|}{\textbf{InterDigital}}               
            & \multicolumn{1}{l|}{\textbf{Qualcomm}}  
\\		
		\endfirsthead
			
			\multicolumn{7}{|l|}%
			{{\bfseries  -- continued from previous page}} \\
			\hline 
		\multicolumn{1}{|l|}{\textbf{Focus Areas}}               
            & \multicolumn{1}{l|}{\textbf{Nokia}}               
            & \multicolumn{1}{l|}{\textbf{Ericsson}}    
            & \multicolumn{1}{l|}{\textbf{Docomo}}               
            & \multicolumn{1}{l|}{\textbf{Samsung}}  
            & \multicolumn{1}{l|}{\textbf{Interdigital}}               
            & \multicolumn{1}{l|}{\textbf{Qualcomm}} 
\\         
			\endhead
			
   \multicolumn{3}{l}{{Continued on next page}} \\ 
			\endfoot

			\endlastfoot
			\hline

\begin{tabular}[|l]{@{}p{3cm}@{}} Ultra-high-speed, low-latency communication  \end{tabular}  &  \multicolumn{1}{c|}{\checkmark} & \multicolumn{1}{c|}{\checkmark}& \multicolumn{1}{c|}{\checkmark}& \multicolumn{1}{c|}{\checkmark}& \multicolumn{1}{c|}{\checkmark}& \multicolumn{1}{c|}{\checkmark} \\
\hline 
\begin{tabular}[|l]{@{}p{3cm}@{}} Massive machine-type \\communications  \end{tabular}  & \multicolumn{1}{c|}{\checkmark} & \multicolumn{1}{c|}{\checkmark}& \multicolumn{1}{c|}{\checkmark}& \multicolumn{1}{c|}{\checkmark}& \multicolumn{1}{c|}{\checkmark}& \multicolumn{1}{c|}{\checkmark} \\
\hline
\begin{tabular}[|l]{@{}p{3cm}@{}} AI and ML  \end{tabular}  & \multicolumn{1}{c|}{\checkmark} & \multicolumn{1}{c|}{\checkmark}& \multicolumn{1}{c|}{\checkmark}& \multicolumn{1}{c|}{\checkmark}& \multicolumn{1}{c|}{\checkmark}& \multicolumn{1}{c|}{\checkmark} \\
\hline
\begin{tabular}[|l]{@{}p{3cm}@{}} Security and privacy  \end{tabular}  & \multicolumn{1}{c|}{\checkmark} & \multicolumn{1}{c|}{\checkmark}& \multicolumn{1}{c|}{\checkmark}& \multicolumn{1}{c|}{\checkmark}& \multicolumn{1}{c|}{\checkmark}& \multicolumn{1}{c|}{\checkmark} \\
\hline
\begin{tabular}[|l]{@{}p{3cm}@{}} Energy efficiency  \end{tabular}  & \multicolumn{1}{c|}{\checkmark}& \multicolumn{1}{c|}{\checkmark}& \multicolumn{1}{c|}{\checkmark}& \multicolumn{1}{c|}{\checkmark}& \multicolumn{1}{c|}{\checkmark}& \multicolumn{1}{c|}{\checkmark} \\
\hline
\begin{tabular}[|l]{@{}p{3cm}@{}} Terahertz \\communications  \end{tabular}  & \multicolumn{1}{c|}{\checkmark} & \multicolumn{1}{c|}{\checkmark}& \multicolumn{1}{c|}{\checkmark}& \multicolumn{1}{c|}{\checkmark}& \multicolumn{1}{c|}{\checkmark}& \\
\hline
\begin{tabular}[|l]{@{}p{3cm}@{}} Network virtualization  \end{tabular}  & \multicolumn{1}{c|}{\checkmark} & \multicolumn{1}{c|}{\checkmark}& \multicolumn{1}{c|}{\checkmark}& \multicolumn{1}{c|}{\checkmark}& \multicolumn{1}{c|}{\checkmark}& \\
\hline
\begin{tabular}[|l]{@{}p{3cm}@{}} Spectrum efficiency  \end{tabular}  & \multicolumn{1}{c|}{\checkmark} & \multicolumn{1}{c|}{\checkmark}& \multicolumn{1}{c|}{\checkmark}& \multicolumn{1}{c|}{\checkmark}& \multicolumn{1}{c|}{\checkmark} & \\
\hline
\begin{tabular}[|l]{@{}p{3cm}@{}} Edge computing \end{tabular}  & \multicolumn{1}{c|}{\checkmark} & \multicolumn{1}{c|}{\checkmark}& \multicolumn{1}{c|}{\checkmark}& \multicolumn{1}{c|}{\checkmark} & & \\
\hline
\begin{tabular}[|l]{@{}p{3cm}@{}} Autonomous networks \end{tabular}  & \multicolumn{1}{c|}{\checkmark} & \multicolumn{1}{c|}{\checkmark}&  &  &  &  \\
\hline

\end{longtable}
\end{small}


\noindent\textbf{Convergence of the physical world and digital/virtual world.} 
    Achieving the convergence of the physical world and digital/virtual world is believed to be an essential path for the 6G communication networks. 
    This integration creates new opportunities and experiences, 
    such as improving network efficiency, 
    enhancing user experiences, 
    increasing world connectivity, 
    facilitating better decision-making, 
    and bringing new business opportunities. 
    However, 
    it gives rise to new challenges related to secure and private data collection/curation.
    In more detail, 
    any data breach in the digital/virtual world could have serious repercussions for the physical world. 
    \begin{itemize}[leftmargin=*]
        \item Increased data collection and storage can compromise personal information in the physical world, 
        such as addresses, payment details, and health records. 
        Individuals may lose control over their personal information and how it is used in the digital/virtual world. 
        \item As more physical devices are connected to the Internet, 
        they become vulnerable to cyberattacks and hacking. 
        Integration of physical systems with digital ones would lead to tampering or attacks on these systems.
    \end{itemize}

\noindent\textbf{Increasing AI computations/functions in 6G.}  With an increased computation power embedded in almost every component of the 6G network, 
    AI algorithms would become prevalent in 6G to perform complex tasks, 
    such as real-time data analysis, 
    autonomous decision-making, 
    and predictive maintenance. 
    In addition, 
    6G networks with improved AI capabilities can optimize network resources,
    reducing energy consumption and costs. 
    It will enable the development of new and innovative use cases for 6G, 
    such as remote surgery, 
    smart cities, 
    and autonomous vehicles. 
    Advanced AI algorithms in 6G networks can help detect and prevent cyber threats and improve overall network security. 
    However, 
    it raises new concerns about the security and privacy of the diversified and pervasive AI models. 
\begin{itemize}[leftmargin=*]
    \item Adversaries may introduce false data or adversarial inputs into AI training, potentially undermining or severely damaging the learning process. They could alter the data, interfere with the model, or change the results. 
    \item Privacy concerns are increasingly prominent in AI systems, particularly due to the centralized nature of AI functions in 6G. The vast datasets and computational resources available in 6G AI learning raise significant data privacy issues.
\end{itemize}

\noindent\textbf{Ubiquitous sensing and high-precision positioning.} Ubiquitous sensing will allow 6G networks to sense and collect data from a wide range of sources.
    In particular, 
    high-precision positioning systems will play a key role in 6G, 
    enabling more accurate positioning and navigation, 
    which will benefit business sectors such as transportation, mining, agriculture, logistics, and autonomous vehicles. 
    6G can leverage ubiquitous sensing and high-precision positioning to create intelligent environments that can understand and respond to their surroundings and provide enhanced connectivity and communication. 
    However, 
    a few issues related to secure and private sensor data generation have to be addressed before the 6G rollout of ubiquitous sensing. 
        \begin{itemize}[leftmargin=*]
            \item Regulating the collection, processing, and usage of sensor data in 6G systems is a grand challenge, 
            especially facing complex and rapidly evolving IoT technologies. 
            \item It is critical to ensure that personal data collected by ubiquitous sensing and high-precision positioning systems is protected from unauthorized access and exploitation. 
        \end{itemize}
As shown in Table~\ref{tab:sp_industry}, {\color{black}some significant security and privacy threats have been identified by the industry}. 
Potential detection and defence methods are also discussed in the table. 

\begin{footnotesize}
\begin{figure*}
    
\begin{longtable}{|p{2cm}|p{4cm}|p{2.5cm}|p{1cm}|p{4cm}|}	
\caption{Security and Privacy threat - 6G industry}
\label{tab:sp_industry}
		\\ \hline             
            \multicolumn{1}{|l|}{\textbf{\begin{tabular}[c]{@{}p{2cm}@{}}{\footnotesize Security and privacy threat to 6G}\end{tabular}}} 
            & \multicolumn{1}{c|}{ {\footnotesize \textbf{Description}}  }
            &\multicolumn{1}{c}{ {\footnotesize \textbf{Consequence}} }
              &\multicolumn{1}{|c|}{\textbf{\begin{tabular}[c|]{@{}p{1.1cm}@{}}{{\footnotesize \textbf{\makecell{Severity\\Likelihood \\Risk}}}}\end{tabular}}}
            & \multicolumn{1}{c|}{\textbf{\begin{tabular}[c|]{@{}p{4cm}@{}}{\footnotesize Potential detection and defence methods}\end{tabular}}} \\
            \hline
		\endfirsthead
			
			\multicolumn{5}{|l|}%
			{{\bfseries \footnotesize -- continued from previous page}} \\
			\hline 
            \multicolumn{1}{|l|}{\textbf{\begin{tabular}[c]{@{}p{2cm}@{}}{Security and privacy threat to 6G}\end{tabular}}} 
            & \multicolumn{1}{|c|}{ {\textbf{Description}}  }             
            &\multicolumn{1}{|c|}{ {\textbf{Consequence}}  }
              &\multicolumn{1}{|c|}{\textbf{\begin{tabular}[c]{@{}p{1cm}@{}}{{\textbf{\makecell{Severity\\ Likelihood \\risk}}}}\end{tabular}}}
            & \multicolumn{1}{c|}{\textbf{\begin{tabular}[c]{@{}p{4cm}@{}}{Potential detection and defence methods}\end{tabular}}} 
			\endhead
			\hline
                \multicolumn{5}{|l|}{{\footnotesize Continued on next page}} \\ 
			\endfoot
			
			\hline
                \hline 
			\endlastfoot

\begin{tabular}[|l]{@{}p{2cm}@{}}\footnotesize Jamming attacks in industrial networks {\color{black}\citep{Nokia_white_paper} } \end{tabular}             
& \begin{tabular}[l]{@{}p{4cm}@{}}\footnotesize External attackers can try to jam networks, making physical security alone inadequate.
 \end{tabular}   
& \begin{tabular}[l]{@{}p{2.5cm}@{}}{\footnotesize This can seriously cripple industrial operations/productions relying on time-sensitive networks.}
 \end{tabular}
& \makecell{\footnotesize High \\ \footnotesize High \\\footnotesize High }
& \begin{tabular}[l]{@{}p{4cm}@{}}{\footnotesize*Energy detection in the radio frequencies 

*Frequency hopping and CDMA technologies to combat jamming signals

*AI-powered real-time analytics can enhance network resilience by adapting to fluctuations in traffic load and radio environments.
}
\end{tabular}  
\\ \hline

\begin{tabular}[|l]{@{}p{2cm}@{}}\footnotesize Security threats in the access control and authorization of 6G sub-networks {\color{black}\citep{Nokia_white_paper}} \end{tabular}             
& \begin{tabular}[l]{@{}p{4cm}@{}}\footnotesize The assets in the 6G sub-networks will not be directly managed by the 6G main networks. Hence, attackers can exploit vulnerabilities in the access control and authorization of 6G sub-networks. 
 \end{tabular}   
& \begin{tabular}[l]{@{}p{2.5cm}@{}}{\footnotesize Unauthorized access to and control of some 6G sub-networks, e.g., body area network (BAN), driverless vehicle system. In some extreme cases, it could endanger human lives. }
 \end{tabular}
& \makecell{ \footnotesize Medium \\ \footnotesize Medium \\ \footnotesize Medium} 
& \begin{tabular}[l]{@{}p{4cm}@{}}{\footnotesize *Some of the critical security functions, e.g., authentication, should be handled by the 6G main network 

*Hardware-based secure environment that provides secure operation of software code and protection of credential
}
\end{tabular}  

\\ \hline

\begin{tabular}[|l]{@{}p{2cm}@{}}\footnotesize Privacy leakage in the creation and transmissions of digital representations of real and virtual objects {\color{black}\citep{Ericsson_white_paper,Nokia_white_paper} } \end{tabular}             
& \begin{tabular}[l]{@{}p{4cm}@{}}\footnotesize When the digital world accurately reflects the physical and biological realms, and new mixed-reality environments blend real and virtual elements, fresh privacy concerns emerge. 
 \end{tabular}   
& \begin{tabular}[l]{@{}p{2.5cm}@{}}{\footnotesize Unintentional leakage of people’s private and sensitive information, e.g., individuals’ demographic, genomic, geographic information }
 \end{tabular}
& \makecell{\footnotesize Medium \\ \footnotesize High \\ \footnotesize High} 
& \begin{tabular}[l]{@{}p{4cm}@{}}{\footnotesize *Privacy-preservation methods, e.g., generalization, perturbation, suppression, etc.
}
 \end{tabular} 
\\ \hline

\begin{tabular}[|l]{@{}p{2cm}@{}}\footnotesize Privacy leakage in the integration of sensors and communication antennas {\color{black}\citep{DOCOMO_white_paper, Nokia_white_paper} }\end{tabular}             
& \begin{tabular}[l]{@{}p{4cm}@{}}\footnotesize 
An adversary could learn individuals’ private information from the sensing data extracted from the communication signals. 
High-frequency bands like mmWave and THz are ideal for fast, high-capacity communication and precise positioning and sensing.
\end{tabular}   
& \begin{tabular}[l]{@{}p{2.5cm}@{}}{\footnotesize Unintentional leakage of people’s private and sensitive information, e.g., human activities, movement trajectories, etc. }
 \end{tabular}
& \makecell{ \footnotesize Medium \\ \footnotesize High \\ \footnotesize High }
& \begin{tabular}[l]{@{}p{4cm}@{}}{\footnotesize *Privacy-preserving provenance of 6G sensing data 
}
\end{tabular}  

\\ \hline

\begin{tabular}[|l]{@{}p{2cm}@{}}\footnotesize Privacy leakage in joint device-network computation {\color{black}\citep{Samsung_white_paper}}  \end{tabular}             
& \begin{tabular}[l]{@{}p{4cm}@{}}\footnotesize 
To address the limitations of mobile device computing power, 6G networks will introduce split computing, utilizing accessible computing resources across the network. Consequently, attackers can infer private information from the data sent from users to different networks.  
\end{tabular}   
& \begin{tabular}[l]{@{}p{2.5cm}@{}}{\footnotesize Unintentional leakage of people’s private and sensitive information, conversation, surroundings, images/videos, etc. }
 \end{tabular}
&  \makecell{ \footnotesize High \\ \footnotesize Low \\ \footnotesize Medium }
& \begin{tabular}[l]{@{}p{4cm}@{}}{\footnotesize *Local privacy preservation methods, e.g., fake backgrounds, identity obfuscation, etc. 

*Advanced encryption technologies 
}
\end{tabular}  
\\ \hline

\begin{tabular}[|l]{@{}p{2cm}@{}}\footnotesize Privacy leakage in distributed learning {\color{black}\citep{qualcomm}} \end{tabular}            
& \begin{tabular}[l]{@{}p{4cm}@{}}\footnotesize 6G networks can initiate distributed learning by calling for users to contribute their data and/or local models for AI training. Attackers can receive/infer private information from the data sent from users to the networks 
\end{tabular}   
& \begin{tabular}[l]{@{}p{2.5cm}@{}}{\footnotesize Unintentional leakage of people’s demographic, identity, health, financial information }
 \end{tabular}
& \makecell{\footnotesize High \\ \footnotesize Low \\ \footnotesize Medium }
& \begin{tabular}[l]{@{}p{4cm}@{}}{\footnotesize *
Ensure transparency by making the system clearly identify how and when it accesses any code, training data, or other resources related to personal information. 

*Privacy-preservation mechanisms to strictly maintain the privacy of user information
}

\end{tabular} 
\\ \hline
\end{longtable}
\end{figure*}    
\end{footnotesize}

\subsubsection{Industrial activities for 6G}

This chapter surveys the efforts of leading companies advancing 6G development. Firms such as Nokia, Ericsson, and Samsung are spearheading research, projects, and partnerships, aiming to revolutionize telecommunications through new technologies, standards, and architectures, collaborating with governments and research institutions to shape the future of global connectivity. 

\noindent\textbf{Nokia.} Nokia is playing a key role in establishing the Horizon Europe Smart Network and Services Joint Undertaking \citep{tsnas}. It has taken a lead role in several research projects aimed at advancing state-of-the-art 6G communication systems. 
On July 11, 2022, Nokia announced its lead role in 6G-ANNA, a German national-funded 6G lighthouse project \citep{ntlg}, collaborating with 29 partners to drive 6G research and standardization. The 6G-ANNA project is focused on developing new technologies and concepts for 6G communication systems, including new frequency bands, advanced modulation and coding schemes, and new network architectures. 
On October 7, 2022, Nokia announced its leadership of the Hexa-X-II project \citep{ntltn}, the second phase of the European 6G flagship initiative, after leading the first Hexa-X project. This new phase expands the number of partners to 44 organizations, tasked with creating the pre-standardized platform and system view for 6G standardization. 
On November 28, 2022, Nokia announced the opening of a new research and development center focused on 5G and future 6G mobile network technology \citep{nton5},  including new radio access technologies, network architecture and design, and software and security solutions, in Amadora, Portugal. 

\noindent\textbf{Ericsson.} Ericsson has a rich history of being an active contributor to standards development in the telecommunications industry. They engage in various research partnerships with government entities from various nations, including the Hexa-X project and the Next G Alliance, which are discussed in Section 3.2. Besides, Ericsson is involved in the Hexa-X-II project again and leads the venture as the technical manager. 

On November 22, 2022, Ericsson declared a ten-year investment plan in the United Kingdom aimed at exploring 6G research and development \citep{eimm}. This investment will see Ericsson pouring millions of British pounds into research fields such as network resilience and security, AI, cognitive networks, and network efficiency. 
In February 2023, Ericsson and KTH announced the launch of DETERMINISTIC6G \citep{nepd}, a research and innovation consortium with a budget of 5.7€ million. The consortium aims to develop networks that can support emerging applications in industrial automation, manufacturing, transport, medicine, and entertainment. Ten partners make up the DETERMINISTIC6G consortium, bringing together mobile network leaders such as Ericsson and Orange, experts in visionary applications and vertical ecosystems like B\&R, IUVO, and SSSA, as well as leading research institutes such as KTH, the University of Stuttgart, and Silicon Austria Labs. The consortium also includes two highly innovative SMEs, Cumucore and Montimage. 

\noindent\textbf{NTT and DOCOMO.} Nippon Telegraph and Telephone Corporation (NTT) and DOCOMO are making significant efforts in their 6G research and development programs, combining NTT's expertise in optical and wireless innovations with DOCOMO's advancements in mobile technology. The two companies have declared that they will work together with Fujitsu, NEC, and Nokia to test cutting-edge mobile communication technologies with the aim of launching 6G services by approximately 2030. Besides, NTT and DOCOMO are involved in various 6G research and development initiatives in collaboration with other companies, such as the Hexa-X project and the 6Genesis project.

Additionally, DOCOMO and Fujitsu are working on a project that has set a goal of creating 6G wireless technology for sub-THz communication utilizing the 100 GHz and 300 GHz frequencies {\color{black}\citep{docomo_fujitsu_6g_2024}}. One key technology in this venture will be the implementation of distributed MIMO. Furthermore, companies are striving to develop high-frequency wireless devices that incorporate compound semiconductors. The companies have been working on this project since 2018, and they have already achieved a number of significant milestones in their research. 

\noindent\textbf{Samsung.} Samsung is taking a leadership role in the development and standardization of 6G. By being named the Chair of ITU-R 6G Vision Group, Samsung is well-positioned to help define the key capabilities, technological developments, and timelines needed to establish 6G standards and bring 6G networks and services to market. 
In 2019, Samsung Research established the Advanced Communications Research Center (ACRC) to advance state-of-the-art communication technologies and contribute to the development of future communication standards {\color{black}\citep{samsung_acrc_2019}}. Also, Samsung Research founded the Next-Generation Communication Research Center, which built upon the foundation of its Standard Research Team, which is the largest unit within Samsung Research and conducts research on 6G technology. The center's research efforts are focused on exploring new technologies and techniques that will be critical to the development of 6G, including advanced antenna systems, mmWave and THz frequency bands, and intelligent networking and computing. 
On October 11th, 2022, Samsung Research announced the creation of a new 6G research group in the UK to concentrate on creating various 6G technologies aimed for implementation in 2030 {\color{black}\citep{samsung_6g_2022}}. This new group in the UK is a notable expansion of Samsung's investment in British telecoms research expertise, forming part of the company's worldwide 6G development initiative that includes numerous research centers abroad.

\noindent\textbf{InterDigital.} InterDigital is actively involved in research and development efforts related to 6G. The company has a number of ongoing initiatives and partnerships focused on advancing the 6G development.
InterDigital is a member of the 6G Flagship program, a Finnish research initiative focused on advancing the development of 6G technology. 
Besides, On November 7, 2022, InterDigital revealed that it has received funding to back five 6G Flagship research projects under Horizon Europe \citep{iafh2022}. These projects, which include 6G-XR, CENTRIC, PREDICT-6G, 6G-BRICKS, and 6G-SHINE, are aimed at promoting technological growth and creating experimental frameworks for the upcoming 6G systems. 

\noindent\textbf{Qualcomm.} Qualcomm is actively involved in research and development efforts related to 6G wireless communication technology. The company is a member of the 6G Flagship program, a Finnish research initiative focused on advancing the development of 6G technology. Qualcomm is also a founding member of the Next G Alliance, a cross-industry group that is working to advance North American leadership in 6G technology.
In addition, 
Qualcomm, along with major wireless industry players such as AT\&T and Samsung, is supporting the University of Texas at Austin in launching a 6G research center called 6G@UT {\color{black}\citep{utnews_6g_2021}}. The center will focus on researching and developing new technologies and applications for the next generation of wireless communication, which is expected to be even faster and more efficient than the current 5G technology. Qualcomm is working with the Wireless Communications and Sensing Laboratory at UC Berkeley on a project focused on developing new technologies and applications for 6G wireless communication. It is collaborating with the Indian Institute of Technology Madras on a project focused on developing  5G/6G technologies.

\subsection{Government projects regarding 6G development}

As a critical infrastructure, governments are involved in 6G study through global initiatives and projects; see Table \ref{tab:gov-proj-apps}.

\begin{table*}[t]
\centering
\caption{Different application scenarios considered in governments
and global 6G initiatives or projects.}
\label{tab:gov-proj-apps}
\begin{footnotesize}
\begin{tabular}{|c|c|ccccccc|c|}
\hline
\multirow{2}{*}{Country} &
\multirow{2}{*}{Project} &
  \multicolumn{8}{c|}{6G Applications} \\ \cline{3-10} 
  & &
  \multicolumn{1}{c|}{\begin{tabular}[c|]{@{}c@{}}UAV \\ mobility\end{tabular}} &
  \multicolumn{1}{c|}{\begin{tabular}[c]{@{}c@{}}Holographic \\Telepresence\end{tabular}} &
  \multicolumn{1}{c|}{\begin{tabular}[c]{@{}c@{}}Autonomous \\Driving\end{tabular}} &
  \multicolumn{1}{c|}{\begin{tabular}[c]{@{}c@{}}Smart \\Grid 2.0\end{tabular}} &
  \multicolumn{1}{c|}{\begin{tabular}[c]{@{}c@{}}Industry 5.0\end{tabular}} &
  \multicolumn{1}{c|}{\begin{tabular}[c|]{@{}c@{}}Digital \\Twin\end{tabular}} &
  \multicolumn{1}{c|}{\begin{tabular}[c|]{@{}c@{}}Intelligent \\Health\end{tabular} }&
  \multicolumn{1}{c|}{\begin{tabular}[c|]{@{}c@{}}Extended\\ Reality\end{tabular} }\\ \hline
  
USA & Next G Alliance &
  \multicolumn{1}{c|}{\checkmark} &
  \multicolumn{1}{c|}{} &
  \multicolumn{1}{c|}{\checkmark} &
  \multicolumn{1}{c|}{\checkmark} &
  \multicolumn{1}{c|}{\checkmark} &
  \multicolumn{1}{c|}{\checkmark} & \checkmark & \checkmark
   \\ \hline
EU & Hexa X &
  \multicolumn{1}{c|}{\checkmark} &
  \multicolumn{1}{c|}{\checkmark} &
  \multicolumn{1}{c|}{} &
  \multicolumn{1}{c|}{} &
  \multicolumn{1}{c|}{\checkmark} &
  \multicolumn{1}{c|}{\checkmark} & \checkmark & 
   \\ \hline
Finland & 6G Flagship &
  \multicolumn{1}{c|}{\checkmark} &
  \multicolumn{1}{c|}{} &
  \multicolumn{1}{c|}{\checkmark} &
  \multicolumn{1}{c|}{\checkmark} &
  \multicolumn{1}{c|}{\checkmark} &
  \multicolumn{1}{c|}{} & \checkmark & \checkmark
   \\ \hline
   Japan & 6G/B5G Strategy &
  \multicolumn{1}{c|}{} &
  \multicolumn{1}{c|}{\checkmark} &
  \multicolumn{1}{c|}{\checkmark} &
  \multicolumn{1}{c|}{} &
  \multicolumn{1}{c|}{\checkmark} &
  \multicolumn{1}{c|}{} & \checkmark & \checkmark
   \\ \hline
   South Korea & MSIT 6G &
  \multicolumn{1}{c|}{\checkmark} &
  \multicolumn{1}{c|}{\checkmark} &
  \multicolumn{1}{c|}{\checkmark} &
  \multicolumn{1}{c|}{} &
  \multicolumn{1}{c|}{} &
  \multicolumn{1}{c|}{\checkmark} & \checkmark & 
   \\ \hline
      Germany & 6GNeXt &
  \multicolumn{1}{c|}{} &
  \multicolumn{1}{c|}{\checkmark} &
  \multicolumn{1}{c|}{\checkmark} &
  \multicolumn{1}{c|}{} &
  \multicolumn{1}{c|}{} &
  \multicolumn{1}{c|}{} &  & \checkmark
   \\ \hline
\end{tabular}%
\end{footnotesize}
\end{table*}

\subsubsection{Next G Alliance}

In early 2021, the Alliance for Telecommunications Industry Solutions (ATIS) launched the Next G Alliance in the United States~\citep{us-nextg}. The primary goal of this initiative is to establish the foundation for 6G leadership within the US and its global counterparts. This program identifies key industries expected to benefit from 6G advancements, including aerospace, agriculture, defense, education, healthcare, manufacturing, media, energy, and transportation, all crucial to the US's future growth and 6G innovation. 

The Next G Alliance is focused on the entire lifecycle of technological commercialization, which includes research and development, production, standardization, and preparing the market for adoption. Among the founding members are major companies such as AT\&T, Google, Intel, Nokia, Samsung, Verizon, and more. The initiative has outlined three initial strategic actions aimed at driving the 6G agenda forward.

\begin{itemize} [leftmargin=*]
    \item Creation of a national 6G roadmap to position North America as a global leader in next-generation communication technologies, focusing on research, development, standardization, and manufacturing, while adapting to the evolving competitive environment.
    \item Unification of the North American technology sector around key priorities to shape public policy and secure funding for 6G advancements.
    \item Formulation of initial strategies and actions for the swift commercialization of Next G technologies across emerging markets and industries, aiming for widespread adoption both domestically and internationally.
\end{itemize}

\subsubsection{Innovate Beyond 5G Program (IB5G)}

The US Department of Defense (DoD) is keenly focused on advancing 5G-to-NextG wireless technologies and supporting concept demonstrations in this area. To further this goal, the DoD has launched the Innovate Beyond 5G program {\color{black}\citep{dod_beyond5g_2022}}, which continues to foster collaboration between the public and private sectors in researching and developing essential 5G-to-NextG wireless technologies. As part of this initiative, IB5G has recently kicked off three key projects aimed at pushing the boundaries of 6G technology research.

\noindent\textbf{Open6G}, is a collaborative effort between industry and academia, specifically focused on initiating 6G systems research in the area of open Radio Access Networks (Open RAN). This initiative centers on researching Open RAN and implementing open-source components of the 5G protocol stack to support the next generation of enhanced 5G applications. Serving as a central hub for developing, testing, and integrating secure advancements, Open6G supports both industry and federal efforts to achieve 6G technology objectives. The project is funded with \$1.77 million from IB5G and is led by Northeastern University’s Kostas Research Institute.

\noindent\textbf{Spectrum Exchange Security and Scalability}, funded by IB5G, is being developed in collaboration with Zylinium Research. This project addresses the growing importance of spectrum-sharing technologies as user demand on wireless networks increases. The research, funded with \$1.64 million from the Office of the Under Secretary of Defense for Research and Engineering (OUSD(R\&E)), led to the creation of a network service appliance called Spectrum Exchange, which is designed to allocate spectrum resources efficiently.

\noindent\textbf{Massive MIMO}, involves a partnership with Nokia Bell Labs. This project, funded with \$3.69 million from OUSD(R\&E)/IB5G through an Open Broad Agency Announcement for Advanced Wireless Communications research, focuses on exploring critical technology components that enable the scalability of MIMO technology across various bands and bandwidths, particularly for DoD-specific use cases.

\subsubsection{Hexa-X}

In 2021, the European Union (EU), {\color{black}in collaboration with Nokia,} launched the Hexa-X project~\citep{hexa-x-report}. This initiative brings together various research institutions and universities to work on the commercialization of cutting-edge technologies, with the goal of establishing the foundation for 6G networks. Hexa-X aims to position itself as a leader in global research and innovation for the next generation of mobile communications. The project is focused on developing the essential tools required to implement 6G networks across Europe. It addresses six key challenges within the 6G landscape: intelligent connectivity, network of networks, sustainability, global service coverage, trustworthiness, and providing extreme user experiences.

Hexa-X is working on formulating strategies to tackle these challenges. AI and machine learning (ML) are being leveraged to enhance communication quality between humans and devices. The project envisions creating a global digital ecosystem that functions as a unified network of networks, making these systems heterogeneous, intelligent, and adaptable. To ensure sustainability, efficient resource management is emphasized, while affordable and practical solutions are being developed to achieve comprehensive global coverage for 6G. The project also prioritizes the creation of highly secure architectures that guarantee data privacy, confidentiality, resilience, and the integrity of communication channels. To push the boundaries of 6G network performance, Hexa-X is focusing on the development of advanced technologies, including AI-powered air interfaces, THz radio access, and network virtualization.

\subsubsection{RISE 6G}

The EU-funded RISE-6G (Reconfigurable Intelligent Sustainable Environments for 6G Wireless Networks) is one of the pioneering 6G projects launched in 2021~\citep{strinati2021}. This project explores innovative solutions that leverage the latest advancements in Reconfigurable Intelligent Surfaces (RIS) to dynamically control radio wave propagation, introducing wireless environments as a service. The consortium includes public, private, and academic participants from several EU countries. RISE-6G focuses on enhancing 6G capabilities with RIS technology by addressing four challenges: developing RIS-assisted signal models, integrating multiple RISs into network architectures, designing use cases for QoS improvements (like energy efficiency and massive capacity), and proposing a prototype for innovation. The project also contributes to standardization efforts.

In addition to the Hexa-X and RISE-6G projects, the EU has funded several other initiatives focused on 6G research. One of these is \textbf{NEW-6G}, coordinated by CEA-Leti, the EU's technology research division. This project aims to bridge the gap between microelectronics and telecommunications, linking networks with equipment and integrating software with hardware. Another significant project is \textbf{AI@EDGE}, which unites European industries, academic institutions, and innovative SMEs to drive progress in AI-for-networks and networks-for-AI beyond 5G systems. This project has two main objectives: (1) creating versatile frameworks for closed-loop network automation that enable flexible, programmable pipelines to develop, deploy, and adapt secure, reusable, and reliable AI/ML models, and (2) building an integrated platform that unifies connectivity and computing to manage resilient, scalable, and secure end-to-end network slices, supporting diverse AI-driven network applications.

\subsubsection{6G Flagship}
The 6G Flagship {\color{black}\citep{6g_flagship_aka_2024}} is an eight-year research initiative funded by the Academy of Finland, focused on creating a “6G-Enabled Wireless Smart Society and Ecosystem.” This project aims to establish new 6G standards tailored for future digital communities. A key objective of the 6G Flagship is to address security and privacy concerns while developing crucial technological components for next-generation mobile networks. The research emphasizes seamless communication among individuals, devices, processes, and objects, raising numerous security and privacy issues that need to be resolved. By tackling these challenges, the project seeks to facilitate the emergence of a highly automated and intelligent society that integrates deeply into various aspects of daily life.

Additionally, the 6G Flagship will conduct extensive pilot programs using a dedicated test network, supported by collaborations between industry leaders and academic institutions. This large-scale research endeavor encompasses three primary domains: wireless communications, computer science and engineering, and electronics and materials. With an estimated total funding of approximately 290 million USD over eight years, the 6G Flagship program is designed to drive significant advancements in these fields.

\subsubsection{MSIT 6G}

South Korea is undertaking an ambitious initiative to roll out 6G networks. In August 2020, the Korean government unveiled a plan to invest 170 million USD in 6G research and development over a five-year period from 2021 to 2026. The Ministry of Science and ICT (MSIT) is spearheading this effort, focusing on research, development, standardization, and eventual deployment of 6G technology~\citep{korea6g-report}. The strategy, titled \lq\lq Leading 6G Era: Imagination into Reality,\rq\rq\ centers on three key areas: the development of next-generation technologies, securing standards and patents, and building the necessary infrastructure.

During the initial phase, MSIT is concentrating on advancing high-risk 6G core technologies through global collaboration. The strategic objectives include achieving a data transmission rate of 1 Tbps, reducing wireless latency to 0.1 ms (with wired latency under 5 ms), extending the connectivity range to 10 km from the ground, incorporating AI throughout the network, and ensuring security is built into the design from the outset. In addition to these goals, the Korean government plans to develop pilot projects for anticipated 6G use cases, such as smart factories, smart cities, and autonomous vehicles. However, the initiative also aims to push beyond these applications, exploring new technologies like 6G satellites to further extend the capabilities and reach of 6G networks.

\subsubsection{6G NeXt Project}

The 6G NeXt (Native Extensions for XR Technologies) project, initiated in late 2022, brings together nine industry and academic partners~\citep{germany6g-report}. Established by Germany’s Federal Ministry of Education and Research (BMBF), this project is part of a broader initiative focused on integrated systems and sub-technologies essential for 6G mobile communications. Deutsche Telekom is leading the 6G NeXt project, with partners including Deutsche Telekom AG, DFKI, Fraunhofer FOKUS, LogicWay GmbH, SeeReal Technologies GmbH, TH Wildau, TU Berlin, TU Ilmenau, Volucap GmbH, Flugplatzgesellschaft Schönhagen mbH, and IDRF e.V.

The 6G NeXt initiative seeks to identify the necessary requirements for 6G networks to enable real-time, highly reliable transmission for advanced XR applications. It will design a modular, scalable, and flexible end-to-end infrastructure, deployed across multiple sites, serving as a testbed for research into complex XR use cases. This new system architecture will be used to implement two innovative and challenging use cases from leading German industries: 

\begin{itemize}[leftmargin=*]
    \item A new anti-collision system for aviation, focusing on drone operations at airports with mixed air traffic. This use case demands ultra-low latency, synchronization of data streams, and the capability for decentralized data processing.
    \item An interactive, real-time 3D holographic video transmission system for video conferencing and monitoring, featuring photorealistic content and accurate 3D depth. This use case requires high upstream and downstream bit rates, along with distributed and intelligent video processing capabilities.
\end{itemize}

\subsubsection{Projects from other countries}
Apart from these projects, countries have initiated 6G research.  

 \noindent\textbf{China:} 
The Chinese government has allocated over \$30 billion for next-generation communication technologies, with a significant focus on 6G. The Ministry of Industry and Information Technology (MIIT) is leading 6G research and development efforts \citep{china6g-report}, and in June 2021, its IMT-2030 Promotion Group published a white paper on 6G \citep{china6g-report}, highlighting capabilities like native intelligence, communication for sensing, digital twins, and enhanced security. Additionally, the China 6G Wireless Technology Task Force is advancing 6G development, and Purple Mountain Laboratories achieved a 6G wireless transmission speed of 206.25 Gbps in a lab setting.

\noindent\textbf{Japan:} 
In 2022, the Japanese government allocated \$450 million to develop a 6G strategy in collaboration with industry and academia \citep{japan6g-report}. In December 2020, Japan established the Beyond 5G Promotion Consortium, which includes major telecom companies and the University of Tokyo
\citep{japan6g-miac}. The consortium promotes global collaboration, hosting the Beyond 5G International Conference in 2021 and signing an agreement with Finland's 6G Flagship program. In the same year, Japan and the US committed \$4.5 billion to jointly advance next-generation communication technologies \citep{fsus}.

\noindent\textbf{UK:} The UK government has announced plans to invest more than £100 million in research and development for next-generation technologies, including 6G, collaborating with vendors like Nokia, Ericsson, and Samsung~\citep{uk-report}. The UK government has earmarked £28 million of this funding for universities in York, Bristol, and Surrey to help design future networks, including 6G. In July 2022, the UK also launched a £3 million partnership with South Korea \citep{uoko}, focusing on Open RAN technology and improving power efficiency, with £1.2 million from the UK government.

\noindent\textbf{Singapore:} 
The Singapore government is increasing 6G research efforts, investing in a 6G lab at the Singapore University of Technology and Design (SUTD) to explore AI, edge computing, and network virtualization \citep{singapore-report}. This is part of a larger \$50 million program under the Future Communications Research and Development Programme (FCP). As part of its "Smart Nation" initiative, Singapore sees 6G as the key to improving citizens' lives. In 2021, Singapore joined the Finnish-led 6G Flagship consortium to contribute to global 6G standards and regulations \citep{singapore-report2}.

\noindent\textbf{India:} 
India is leading the 6G development in Southeast Asia. On November 1, 2021, the Department of Telecommunications (DoT) formed the Technology Innovation Group on 6G (TIG-6G) with representatives from various sectors to define India's 6G vision and objectives \citep{india-report}. Additionally, TIG-6G also created a roadmap and action plans to guide the country's 6G journey. In January 2022, the DoT established a 6G Technologies Division to focus on advancing 6G, Quantum Communication, Passive Optical Networks, Green Telecom, and critical communications, such as Public Protection and Disaster Relief (PPDR).

\section{Standardization Activities} \label{sa}

This section outlines the standardization processes in the ICT sector, focusing on the structure and activities of the International Telecommunication Union (ITU). It also covers the standardization of mobile communication systems from 3G to 6G and highlights contributions from key organizations like 3GPP, GSMA, and the O-RAN Alliance.

\smallskip 
\noindent 
\textbf{ITU. }
The International Telecommunication Union (ITU)~\citep{aitui} is a United Nations agency~\citep{tun} that specializes in matters related to information and communication technologies (ICTs). ITU's activities are organized into three primary Sectors: the Radiocommunication Sector (ITU-R), the Telecommunication Standardization Sector (ITU-T), and the Telecommunication Development Sector (ITU-D). These Sectors carry out their work through various conferences and meetings. ITU-R and ITU-T are responsible for developing standards, known as ITU-R Recommendations and ITU-T Recommendations, respectively. ITU-R Recommendations focus on services that utilize radio frequency technologies, while ITU-T Recommendations address other ICT-related issues. ITU-D is dedicated to bridging the digital divide between developed and developing nations.

The work within each ITU Sector is further divided into Study Groups. ITU-R, for example, is currently organized into five Study Groups, covering areas such as Spectrum Management, Radiowave Propagation, Satellite Services, Terrestrial Services, and Broadcasting and Science Services. The term International Mobile Telecommunications (IMT) is used by ITU to refer to broadband mobile systems, which include 3G, 4G, and 5G technologies. While 3G, 4G, and 5G are widely recognized terms, ITU uses the formal designations IMT-2000, IMT-Advanced, and IMT-2020, as established in ITU-R Resolution 56. The upcoming 6G technology is expected to be formally designated as IMT-2030. Work related to IMT is managed by the Terrestrial Services Study Group (SG5).
As of February 2023, ITU-R Working Party 5D is in the process of developing the framework and overall objectives for IMT-2030, with the goal of completing this work by the end of 2023.
In ITU-T, activities are divided among 11 Study Groups. Study Group 17 (SG17) is responsible for coordinating work related to security. According to the executive summary~\citep{es} of the August/September 2022 ITU-T SG17 meeting, several key topics have been identified, including security for intelligent transport systems, updated TTCN-3 standards, 5G security, cloud security, distributed ledger technology (DLT)-based security services, and IoT security. However, the role that ITU-T SG17 will play in the development of 6G standards remains unclear.

\smallskip 
\noindent 
\textbf{3GPP. }
The 3rd Generation Partnership Project (3GPP) develops Technical Specifications that have been adopted by the ITU-R as standards for 3G (IMT-2000), 4G (IMT-Advanced), and 5G (IMT-2020). The structure of 3GPP includes a Project Coordination Group (PCG) and several Technical Specification Groups (TSGs), which further establish Working Groups (WGs). The PCG is the highest decision-making authority within 3GPP, meeting every six months to finalize the adoption of TSG Work Items, ratify election results, and confirm resource commitments to 3GPP. As of December 2022, various Working Groups under the Technical Specification Group Service and System Aspects (TSG-SA) are active, with TSG-SA3 specifically focusing on Security and Privacy.

3GPP operates using a system of parallel Releases, which provides developers with a stable platform to implement new or significantly enhanced features at specific points in time, allowing for further feature additions in future Releases. A detailed overview of 3GPP Releases can be found in \citep{3gpp_releases}. Release 15 (Rel-15) was submitted to ITU-R as a candidate for 5G (IMT-2020). Currently, Release 18 (Rel-18) is known as 5G Advanced, and subsequent Releases, such as Release 21 (Rel-21), may be considered candidates for 6G (IMT-2030). TSG-SA3 is presently working on two main tasks: 1) addressing known issues from previously published Releases (up to Release 17) and 2) tackling study items for the upcoming Release (currently Release 18).

\smallskip 
\noindent 
\textbf{GSMA. }
GSM Association (GSMA) consists of approximately 750 MNOs and 400 associate members active in the mobile ecosystem. GSMA is said to deliver for its members across three broad pillars as follows \citep{utmear}:

\begin{itemize}[leftmargin=*]

\item {\textbf Connectivity for Good} focuses on collaborating with members, governments, and civil society to promote positive policy outcomes and efficient spectrum usage. This pillar aims to drive digital innovation to address global inequalities and address significant societal challenges, including digital inclusion, climate change, and sustainability.

\item {\textbf Industry Services and Solutions} supports the technology and interoperability that power the mobile industry. Through various projects, working groups, and promotional efforts, GSMA helps the industry concentrate on key areas like 5G, Mobile IoT, fraud prevention, and security. GSMA also provides technical services that offer tools, data, and resources to enhance the efficiency and reliability of mobile experiences for users.

\item {\textbf Outreach} organizes and hosts the annual Mobile World Congress~\citep{tfmmb}, bringing together stakeholders from across the mobile ecosystem. Additionally, GSMA provides breaking news, insights, and expert analysis through platforms like Mobile World Live~\citep{mwl} and GSMA Intelligence~\citep{gi}.

\end{itemize}

In the realm of security and privacy, GSMA offers the GSMA Network Equipment Security Assurance Scheme (NESAS)~\citep{gsmanes}, which serves as a global security assurance framework. This framework is designed to audit and test network equipment vendors and their products against a defined security baseline. NESAS consists of two primary components: the ``Product Development \& Lifecycle Management Process Audit'' and the ``Product \& Evidence Evaluation.'' These evaluations are conducted by authorized GSMA NESAS Test Laboratories in line with the 3GPP Security Assurance Specifications. Currently, NESAS is aligned with 5G standards, but it is anticipated that the scheme will expand to include 6G as development in that area advances.

\smallskip 
\noindent 
\textbf{O-RAN Alliance. } 
The O-RAN Alliance was founded in February 2018 by five mobile network operators—AT\&T, China Mobile, Deutsche Telekom, NTT DOCOMO, and Orange—with the goal of enhancing interoperability among network equipment components to drive market competition and foster innovation. By December 2021, the O-RAN Alliance had grown to include over 160 companies, including 24 mobile operators spanning four continents~\citep{0323}. The O-RAN Alliance focuses on developing specifications that complement those of 3GPP, with the aim of realizing Open RAN. While the adoption of Open RAN is expected to offer greater flexibility, improved interoperability, and lower capital and operational costs, a risk assessment of its current implementation has identified numerous high-risk threats~\citep{0324}. Recognizing the importance of addressing these risks, the Australian Department of Home Affairs and the U.S. Department of Commerce issued a joint statement in September 2022~\citep{so5ris}, agreeing to collaborate on advancing the technical security of O-RAN.

\section{Open research issues}\label{ori}
In this section, {\color{black}we identify several promising research issues that have not been well studied for 6G networks. }

\smallskip 
\noindent\textbf{Post-Quantum Cryptographic Protocols.}
As quantum computing progresses, traditional cryptographic algorithms become vulnerable to quantum attacks, threatening the security of 6G networks. Research may focus on developing and implementing post-quantum cryptographic protocols, such as those recommended by NIST, to ensure the resilience of 6G communications against future quantum threats. This involves selecting algorithms that offer robust security while addressing performance and efficiency concerns inherent to post-quantum cryptography. 

\smallskip 
\noindent\textbf{Network Anomaly Detection.} With the increasing complexity and data flow in 6G networks, leveraging AI-driven approaches like federated learning for real-time network anomaly detection becomes crucial. These systems must be capable of identifying unusual patterns and potential cybersecurity threats swiftly to maintain network integrity. Research should focus on enhancing these AI models' accuracy and reliability in detecting and responding to anomalies across the highly dynamic and distributed 6G environment. 

\smallskip 
\noindent\textbf{Regulatory Compliance.} Ensuring compliance with data protection regulations like GDPR is a significant challenge in the context of 6G, particularly regarding the movement and storage of data across different jurisdictions. Research should focus on developing mechanisms that enforce regulatory compliance seamlessly, adapting to varying legal requirements while safeguarding user data. This includes creating adaptable virtual on-board units (vOBUs) that can manage data protection in diverse legal environments.

\smallskip 
\noindent\textbf{Security and Privacy in MEC.} MEC introduces new security and privacy challenges due to its proximity to end users and the diverse applications it supports. Research should focus on developing security measures that ensure the isolation of sensitive workloads and protect user data in MEC environments. This is critical for maintaining the security and privacy of applications, such as virtual reality and extended reality in 6G networks. 

\smallskip 
\noindent\textbf{Energy-Efficient Security Solutions.} As 6G networks strive for high energy efficiency, security solutions must also be optimized for low energy consumption. Research should explore lightweight cryptographic protocols and energy-efficient security mechanisms that do not compromise the performance or security of the network. This includes developing efficient encryption algorithms and security frameworks that align with the energy constraints of 6G. 

\smallskip 
\noindent\textbf{Risk Assessment Frameworks for 6G.} Existing risk assessment frameworks may not adequately address the complexities of 6G networks, necessitating the development of new models that focus on data privacy and confidentiality. Research should aim to create tailored risk assessment frameworks that can evaluate and mitigate risks specific to 6G environments, ensuring comprehensive protection of user data and network integrity. These frameworks must be capable of identifying potential privacy risks, assessing their impact, and providing actionable mitigation strategies. 

\section{Conclusion}\label{conclusion}
Wireless communication has evolved significantly, culminating in the emergence of 6G, which offers enhanced data speeds, near-zero latency, and increased capacity for a variety of connected devices.
However, the increased complexity of 6G networks introduces new challenges, particularly concerning security and privacy. The integration of diverse IoT devices, AI, and edge computing raises the risk of data breaches and unauthorized access, necessitating robust security measures and privacy technologies. 
This survey provides an overview of 6G's development, delving into their architecture, protocols, security, privacy, and standardization. It identifies existing and emerging security and privacy risks, including vulnerabilities inherited from 5G, potential attacks on 6G protocols, and privacy concerns from new 6G applications. It also discusses mitigation strategies along with efforts by industry, government, and standardization bodies. Additionally, we identify several research directions for future exploration.


\begin{acks}
This research paper is conducted under the 6G Security Research
and Development Project, as led by the Commonwealth
Scientific and Industrial Research Organisation (CSIRO)
through funding appropriated by the Australian Government’s
Department of Home Affairs. This paper does not reflect any
Australian Government policy position. For more information
regarding this Project, please refer to
https://research.
csiro.au/6gsecurity/.
\end{acks}

\bibliographystyle{ACM-Reference-Format}
\bibliography{sample-base}


 




\vfill

\end{document}